\documentclass[preprint,12pt,authoryear]{elsarticle}

\usepackage{amssymb}
\usepackage{lineno}
\usepackage{booktabs}
\usepackage{bm}
\usepackage{multirow}
\usepackage[table,xcdraw]{xcolor}
\usepackage{xr-hyper}
\usepackage{hyperref}
\usepackage{gensymb}
\usepackage{xcolor}
\usepackage{soul}
\usepackage[normalem]{ulem}
\usepackage{booktabs}
\usepackage{multirow}

\journal{Energy}
\begin{document}
\begin{frontmatter}

\title{Novel strategies of Ensemble Model Output Statistics (EMOS) for calibrating wind speed/power forecasts}

\author[DICCA]{Gabriele Casciaro} \ead{gabriele.casciaro@edu.unige.it} 

\author[inst3,INFN]{Francesco Ferrari} \ead{francesco.ferrari2@unimi.it}

\author[DICCA,INFN]{Mattia Cavaiola} \ead{mattia.cavaiolao@edu.unige.it}

\author[DICCA,INFN]{Andrea Mazzino\corref{cor}} \ead{andrea.mazzino@unige.it}

\affiliation[DICCA]{organization={DICCA, Department of Civil, Chemical and Environmental Engineering. University of Genoa},
            addressline={Via Montallegro 1}, 
            city={Genoa},
            postcode={16145}, 
            state={Genoa},
            country={Italy}}
\affiliation[INFN]{
            organization={INFN, National Institute of Nuclear Physics, Genoa section},            addressline={Via Dodecaneso 33}, 
            city={Genoa},
            postcode={16146}, 
            state={Genoa},
            country={Italy}}
\affiliation[inst3]{organization={Ardito Desio Department of Earth Sciences, University of Milan},
            addressline={Via L. Mangiagalli 34}, 
            city={Milan},
            postcode={20133}, 
            state={Milan},
            country={Italy}}
\cortext[cor]{Corresponding author}

\begin{abstract}

The issue of the accuracy of wind speed/power forecasts is becoming more and more important as wind power production continues to increase year after year. Having accurate forecasts for the energy market clashes with intrinsic difficulties of wind forecasts  due to, e.g., the coarse resolution of Numerical Weather Prediction models. Here, we propose a novel Ensemble Model Output Statistics (EMOS) which accounts for nonlinear relationships between predictands and both predictors and other weather observables used as conditioning variables. The strategy is computationally cheap and easy-to-implement with respect to other more complex strategies dealing with nonlinear regressions. Our novel strategy is assessed in a systematic way to quantify its added value with respect to ordinary, linear, EMOS strategies. Wind speed/power forecasts over Italy from the Ensemble Prediction System (EPS) in use at the European Centre for Medium-Range Weather Forecasts (ECMWF) are considered for this purpose.  The calibrations are based on the use of past wind speed measurements collected by 69 SYNOP stations over Italy in the years 2018 and 2019. Our results show the key role played by conditioning variables to disentangle the model error thus allowing a net improvement of the calibration with respect to ordinary EMOS strategies.
Finally, we have quantified the impact of calibrated wind forecasts on the wind power forecasts finding results of interest for the renewable energy market.

\end{abstract}

\begin{graphicalabstract}

\includegraphics[width=\textwidth]{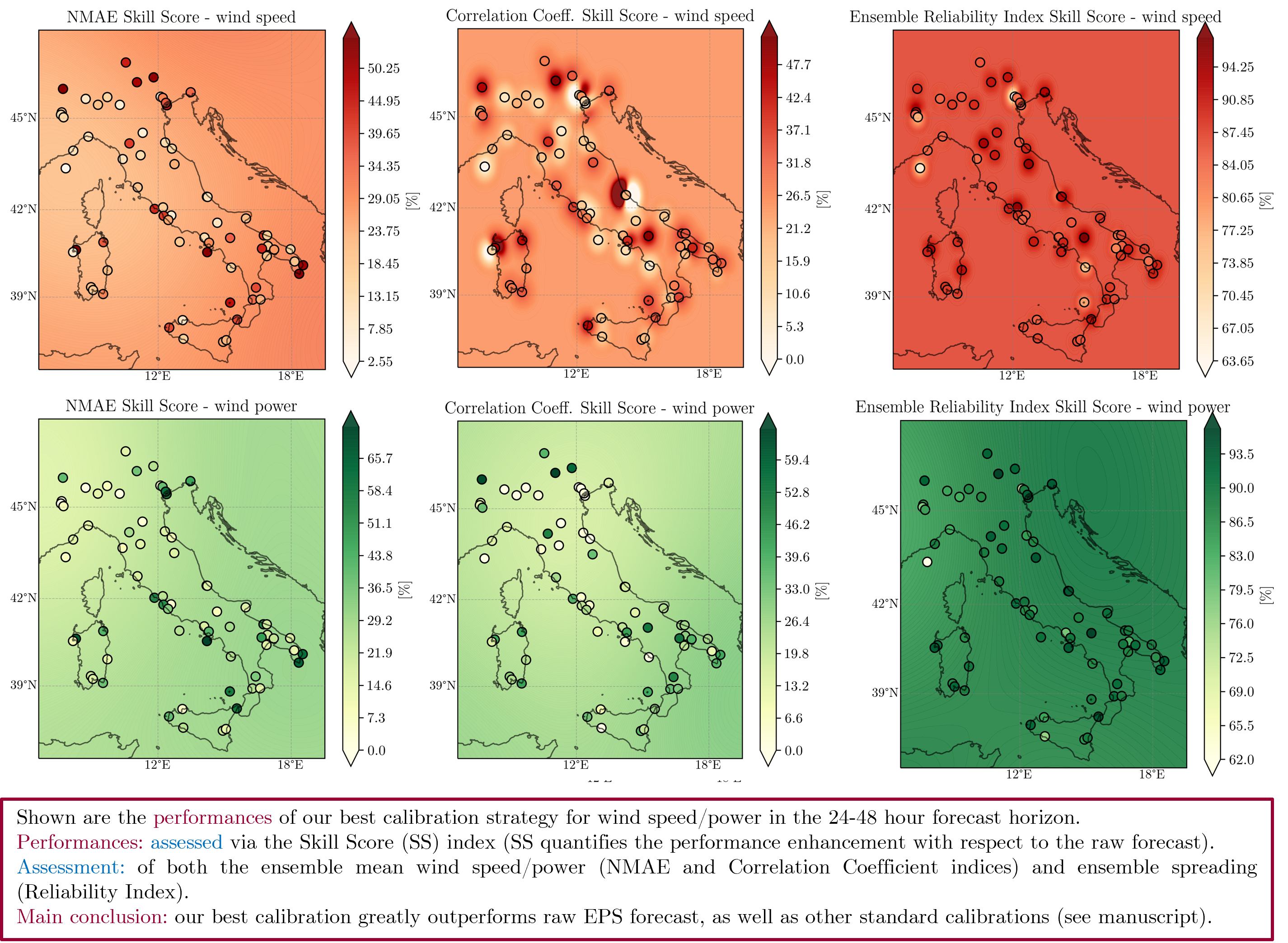}

\end{graphicalabstract}

\begin{highlights}

\item From nonhomogeneous linear regressions to nonhomogeneous nonlinear regressions: a novel easy-to-implement EMOS approach.

\item Nonlinear features are easily and economically accounted for in terms of suitable conditioning meteorological variables.

\item Nonlinear features greatly improve the ordinary EMOS performances.

\item Our best calibration for the wind speed provides a clear added value for the wind power forecast as well.

\end{highlights}


\end{frontmatter}

\section{Introduction}
\label{Introduction}
Wind industry at global level has obtained in 2020 its best performance with a year-over-year growth of 53 $\%$ thus reaching a global cumulative wind power capacity up to 743 GW. This trend is expected to continue and before 2025 wind industry is expected to exceed 1 TW in global cumulative installations of onshore and offshore wind \citep{council2021gwec}.\\
The growing importance of wind industry is accompanied by an increasing contribution of wind power in power systems (the so-called penetration), a fact calling for a new definition of a modern, more complex, concept of flexibility in power systems (see \cite{impram2020challenges} and references therein). Wind is indeed highly intermittent in space and in time and thus very challenging to predict, even for the shortest look-ahead forecast horizons of interest for the energy industry. Because of the nonlinear (cubic) relationship between wind speed and wind power, forecasting with accuracy this latter is even more challenging than wind speed.\\
It then follows from the above considerations that detailed schedule plans and reserve capacity must be properly set by  power system regulators according to a new definition of flexibility \citep{impram2020challenges}.\\

In order to efficiently deal with the continuous increase of the wind power production, the issue on the accuracy of wind forecasts becomes of paramount importance.  Accordingly, several wind speed forecasting methods have been reported in the literature over the past few years. For a comprehensive review readers are referred to \cite{soman2010review}. Our focus here is on wind forecasts with look-ahead time up to 2 days, belonging to the so-called long-term forecasting, from 1 day to 1 week ahead according to \cite{soman2010review}. For such forecast time horizons, physical methods based on Numerical Weather Prediction (NWP) models appear the best strategy, especially when accompanied by statistical methods (which, used alone, perform well for short-term wind predictions) to train the NWPs on the local conditions (orographic and/or dynamical and/or seasonal) via suitable, sufficiently long, training set of forecast-observation pairs.\\
However, to efficiently deal with the issue of wind penetration, power system regulators call for additional information on the uncertainty of wind speed/power forecasts \citep{giebel2011state, monteiro2009quick} when scheduling power reserves. The same request comes from traders for  marketing wind energy. The answer to the end-user stimuli is to associate a deterministic forecast with a probabilistic one \citep{zhang2014review}.
The strong potential of uncertainty forecasting, to go along with point forecasting, for wind power has been clearly shown both exploiting
numerical-weather-prediction-based  methods (see, e.g., \cite{taylor2009wind}) and statistical-based methods (see, e.g., \cite{zhang2016direct,chai2019conditional}).
End-users have in this way information about prediction uncertainty together with a single forecast power value (corresponding, e.g., to the most probable prediction) for each forecast time horizon.\\
In the present paper we present novel calibration strategies both to correct `raw' ensemble weather forecasts and to transform past observed data into accurate wind speed forecasts usable for the renewable energy industry when forecasts from NWP models are not available. As far as the probabilistic NWP model is concerned,
the ECMWF-EPS (Ensemble Prediction System in use at the European Centre for Medium-Range Weather Forecasts) is considered.
We propose a statistical calibration of its ensemble wind speed members based on the use of past wind speed measurements collected by 69 SYNOP stations over Italy (see Fig.\ 1) in the years 2018 and 2019.
Our strategies can be applied to any ensemble forecasting system operative in other weather centres as, e.g., M\'eteo France (the 35-member ARPEGE1-EPS, described by \cite{descamps2015pearp}), the Hungarian Meteorological Service (the 11-member AROME-EPS, described by \cite{javorne2020way}), among others.\\
The choice of the Italian territory is motivated from the fact that Italy presents a huge variety in its orographic shape, ranging from alpine regions, with station elevations up to 3500 meters, to flat areas as, e.g., the Padana valley. Moreover, the mutual interaction between land and sea circulations makes wind prediction a very hard problem.
Because of their uniformly distributed character, SYNOP data over Italy are thus an ideal framework to assess the skill of all our proposed calibrations. The conclusions we will draw are expected to hold, {\it a fortiori}, for regions in the world where weather forecasts are easier due to geographic/orographic/dynamical reasons.\\
Our novel calibrations are evolutions of the well-known Ensemble Model Output Statistics (EMOS) (see, e.g., \cite{gneiting2005calibrated, wilks2007comparison, delle2013probabilistic,wilks2018univariate} among the others) where the used predictive probability density function involved in the standard EMOS strategy is now made conditional on several meteorological observables expected to be useful to disentangle the EPS-based forecast error.
Conditioning variables will allow us to easily insert in the calibration strategy a dependence of EMOS free parameters on the state of the atmosphere, thus assigning to our strategy a dynamical character. One can in this way easily and economically capture nonlinear relationships between
predictands and both predictors and other weather observables used as conditioning variables without having to specify appropriate link functions.
Thanks to the conditioning variables, the standard EMOS
  (i.e.\ a homogeneous linear regression) transforms into a nonhomogeneous nonlinear regression.
Other techniques have been recently  designed to incorporate nonlinear relationships between arbitrary predictor variables and forecast distribution parameters to be determined in a data-driven way, and are based on neural networks (\cite{rasp2018neural,baran2021calibration}). Our approach thus shares with more complex strategies similar objectives  while keeping complexity from a technical point of view at minimum.

We will also deal with other important issues related to the coarse character of the EPS computational grid (about 18 km over Italy) impacting the wind forecasts in a given location (e.g., associated to a wind turbine) at a given height (e.g., the hub height). This issue will be incorporated in a further evolution of the standard EMOS strategy. All steps defining our different calibration strategies will be assessed  in terms of proper statistical indices. In particular, the Skill Score (SS) will be extensively used throughout the paper to quantify how an improved calibration outperforms less elaborated calibrations and/or simpler forecasts based on persistence/climatology.
The superiority of our proposed best calibration with respect to a  standard EMOS calibration will clearly emerge from our analysis. The added value of our strategy is brought by nonlinear features economically inserted into our model via conditioning variables.\\
As far as observation-driven forecasts are concerned, we propose a simple calibration which makes persistence interesting in situations where forecasts  based on NWPs are not available.\\
Our paper does not deal with the sole issue of calibration for the wind speed but also addresses the question on the added value of a calibrated wind forecast on the wind power forecast. The cubic relationship between wind speed and  wind power tends indeed to magnify the relative errors associated to the wind forecast. It is worth emphasizing that our performance assessment is not restricted to the sole mean properties of the EPS but is also aimed at quantifying the improvement of our calibrations in relation to the wind speed/power EPS-based probability density function as a whole. This step is crucial for the ultimate goal of any ensemble forecast for the wind energy market of obtaining an accurate estimation of the uncertainties associated to the mean EPS wind/wind power forecasts.\\
A final note on the general character of our results is worth emphasizing. Although our calibrations have been tested on a large number of SYNOP stations over Italy, all the proposed calibrations can be implemented in any location in the world, on the ground as well as on the sea, at the ground level as well as at higher elevations. What one needs is a sufficiently long record of past
observation-forecast pairs
of the weather observable subject to calibration (here the wind speed) together with a past record of forecast of observables to be used for conditioning.

The paper is organized as follow. Sec.~2 is devoted to describing our ensemble forecasts and our SYNOP observations over Italy. In Sec.~3 we introduce the concept of calibration for the wind speed via the so-called Ensemble Model Output Statistics (EMOS). We will quickly summarize the known relevant literature on the subject and present our contributions on how to further generalize current known strategies. In Sec.~4 the relevant statistical indices to assess our calibrations will be presented. Sec.~5 presents our results starting from a refined definitions of persistence based on a suitable calibration procedure. We will then move to discuss the results on our different proposed calibrations highlighting the level of improvement for each of them in terms of a comparative analysis carried out via suitable statistical indices. Sec.~6 addresses the issue of the impact of a calibrated wind forecast on the wind power forecast. Conclusions and perspectives are finally drawn in the final section.


\section{Wind data}
\label{Wind data}
\subsection{Observed data: SYNOP stations}
\label{Observed data: SYNOP stations}
\begin{figure}[h!]
\includegraphics[width=\textwidth]{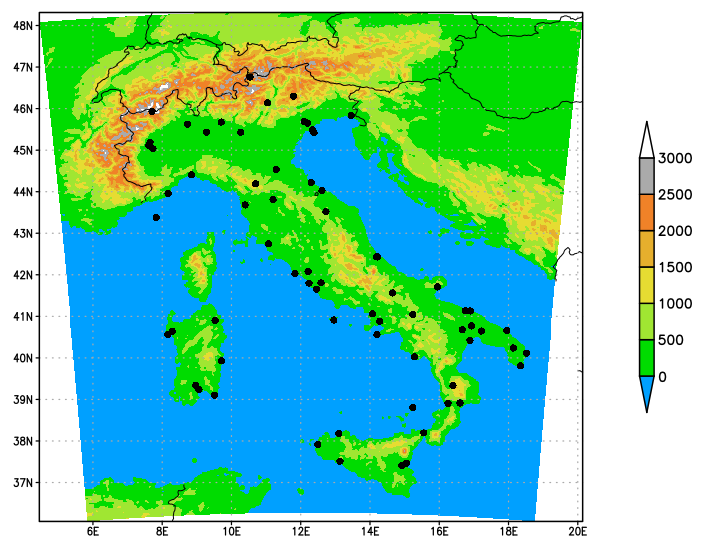}
\caption{Location of SYNOP stations throughout the Italian peninsula used in the present study. Colors are coded according to the orography elevation.}
\label{fig:stazioni_synop_italia}
\end{figure}

For 69 sites around Italy, SYNOP meteorological readings were acquired from year 2015 to year 2020.
The SYNOP anemometers records the wind speed, measured in knots, as an average over 10 minutes according to ICAO specifications at the nominal measurement height of 10 meters a.g.l. \citep{icao2007meteorological}.\\
The time intervals between successive observations may vary between different SYNOP stations, and the hours of operation of meteorological stations are dependent on the specific regime of the anemometer. Smaller anemometers can run from 4 to 18 hours per day, whereas larger anemometers often run 24 hours per day, with an acquisition frequency of one or three hours.\\
As one can see from Figure\ \ref{fig:stazioni_synop_italia}, the stations are arranged rather uniformly across the peninsula.\\

\subsection{Forecast data: the ECMWF Ensemble Prediction System (EPS)}
\label{Forecast data: the ECMWF Ensemble Prediction System (EPS)}

The ECMWF Ensemble Prediction System (EPS) is composed of 51 members, including a control forecast with no perturbations and 50 forecasts created by adding small perturbations to the best-known initial condition, following a mathematical formulation based on singular vector decomposition and stochastic parameterization representing the model uncertainties \citep{buizza1995optimal, leutbecher2008ensemble}.\\
The EPS considered in the present study has spectral triangular truncation with a cubic-octahedral grid Tco639 and 91 layers with a top of the atmosphere of 0.01 hPa \citep{buizza2018ensemble}; it has a resolution of about 18 km \citep{persson2001user}.
The EPS run we have considered is the one created at the 00 UTC of which we have analyzed the sole first 48-h look-ahead time.\\
The mean and variance of the ensemble usually have a good correlation with observations and the actual model uncertainty, respectively, but often tend to underestimate and to be underdispersive with respect to the true observations and uncertainty of the model, respectively \citep{molteni1996ecmwf, montani2019performance}. For these reasons, a calibration must be carried out downstream of this forecast.


\section{The issue of calibration: the Ensemble Model Output Statistic (EMOS) approach}
\label{The issue of calibration: the Ensemble Model Output Statistic (EMOS) approach}
The Ensemble Model Output Statistic (EMOS) is an easy to implement statistical post-processing technique proposed by \cite{gneiting2005calibrated} that allows the calibration of an ensemble forecast as the EPS.
EMOS is an evolution of the multiple linear regression or Model Output Statistic (MOS)  \citep{glahn1972use} which is used commonly to calibrate deterministic forecasts.\\

\subsection{The basic EMOS}
\label{The basic EMOS}
Let $X_1, \cdots, X_K$ indicate $K$ ensemble member forecasts of a univariate continuous, positive-defined, variable $Y$, the weather quantity of interest such as wind speed at a specific locations and look-ahead time.\\
EMOS approach employs  parametric distribution of the general form:

\begin{equation}
Y \mid X_1, \cdots, X_K \thicksim f(Y \mid X_1, \cdots, X_K)
\label{eq:MOS}
\end{equation}
where the left-hand-side means that the distribution is conditional, i.e. the ensemble members are given.\\
Note that if $Y$ is the wind speed, the Weibull distribution is not suitable for $f$ because it represents the unconditional (marginal) distribution of wind speed \citep{thorarinsdottir2010probabilistic}.\\
In way of example, \cite{gneiting2006calibrated} proposed for $f$ the truncated normal distribution (TN) to model the conditional distribution for the wind speed.
The log-normal (LN) distribution was proposed  by \cite{baran2015log} where
it has been found that the TN-LN mixture model outperforms the traditional TN.
The gamma distribution is instead used here following \cite{scheuerer2015probabilistic}
Extensive discussions and comparisons among the different strategies are reported by \cite{wilks2018univariate}. \\Coming back to the predictive gamma Probability Density Function (PDF) used here, it is denoted as

\begin{equation}
\mathcal{G}(\mu, \sigma^2)
\label{eq:EMOS_distr}
\end{equation}
with mean, $\mu$, and variance, $\sigma^2$, given by

\begin{equation}
\mu = a+b_1 X_1 + ... + b_K X_K
\label{eq:EMOS_mu}
\end{equation}

\begin{equation}
\sigma^2 = c+dS^2 .
\label{eq:EMOS_sigma}
\end{equation}
here, $a, b_1, \cdots , b_K, c, d$ are the non-negative EMOS coefficients, and $S^2$ is the ensemble spread defined as the variance of the EPS.\\
The gamma distribution is a two-parameter family of continuous probability distributions \citep{wilks2011statistical}. Among the different possible parameterizations, here we use the shape parameter $k$ and the scale parameter $\theta$. In plain words, $k = \mu^2 / \sigma^2$ and $\theta = \sigma^2 / \mu$.\\
In order to determine the EMOS coefficients, \cite{gneiting2005calibrated} proposed a strategy based on the minimization of the Continuous Ranked Probability Score (CRPS) \citep{hersbach2000decomposition}. This latter is defined as:

\begin{equation}
CRPS(F,Y) = \int_{-\infty}^{\infty} [F(t) - H(t-Y)]^2 dt
\label{eq:crps_generale}
\end{equation}
where $F$ is the cumulative probability of $\mathcal{G}$, $Y$ is the observation and $H$ is the Heaviside function and takes the value 0 when $t < Y$ and the value 1 otherwise.\\
For the gamma distribution a closed form for the CRPS has been obtained by \cite{scheuerer2015probabilistic}, facilitating the minimization procedure. For an observation-forecast pair ($Y, \mathbf{X}$) it reads:
\begin{equation}
crps =  Y\left[2P\left(k, \frac{Y}{\theta}\right)-1\right]-k \theta\left[2P\left(k+1, \frac{Y}{\theta}\right)-1\right]-\frac{\theta}{\beta\left(\frac{1}{2},k\right)}
\label{eq:crps}
\end{equation}
where $Y$ is the observation, $P$ the incomplete gamma function \citep{abramowitz1948handbook} and $\beta$ the beta function. The forecast vector $\mathbf{X} = (X_1, \cdots, X_K)$ enters in (\ref{eq:crps}) via $k$ and $\theta$.\\
In a training set where both observations and forecasts are available, the quantity to be minimized is:

\begin{equation}
CRPS = \frac{1}{N}\sum_{i = 1}^N crps(\mathbf{X}_i, Y_i)
\label{eq:CRPS_tot}
\end{equation}
where $i$ stands for the i-th observation-forecast pair
and $N$ is the total number of pairs in the training set.\\
This strategy satisfies all criteria proposed by \cite{gneiting2007probabilistic} for evaluating predictive performance, which are based on the paradigm of maximizing the sharpness of the predictive distributions that are subject to calibration where sharpness is a property of forecasts that refers to the concentration of predictive distributions (for more details see \cite{gneiting2007probabilistic}).\\
We dub the ordinary EMOS strategy as EMOS$_0$.

\subsection{Evolved EMOS strategies}
\label{The evolved EMOS}

Three main aspects characterize a standard EMOS approach:

\begin{enumerate}
\item the training period is typically a rolling window of 40 days before the day of the forecast \citep{gneiting2005calibrated};
\item forecasts in (\ref{eq:crps}) are usually referred to the grid points closest to the observations;
\item the predictive distribution is conditional to the sole ensemble observables one wants to forecast \citep{thorarinsdottir2010probabilistic}.
\end{enumerate}
Here, we relax all three points and propose multiple variants of the original EMOS strategy.
In particular, relaxing point 3.,  nonlinear relationships
  between predictands and both predictors and other weather observables used as conditioning variables  can be easily
  accounted for as shown in  the next section.

\subsubsection{The EMOS$_+$ strategy}
\label{The EMOS_+ strategy}

The idea of EMOS$_+$ is to consider parametric distributions conditional on a larger set of observables other than the weather quantity of interest (here the wind). The underlying idea is that forecast errors may depend on the specific meteorological conditions occurring at the forecast time in the specific location. \\
The following simple example helps to understand the basic idea behind conditioning variables.
Let us consider a given SYNOP station and that east of the station there is a hill not resolved by the EPS model (e.g., a hill having a typical size smaller than the EPS resolution). We thus expect larger forecast errors for winds blowing from east than for winds blowing, e.g., from north. This implies a different structure of the model error depending on the direction of the wind, a fact that implies different values of the parameters entering the 
EMOS predictive distribution, the values of which depend on the wind direction.  A similar way of reasoning can be extended to other meteorological variables.\\
Moving to a more quantitative description, the EMOS strategy in its standard form modifies in:
\begin{equation}
Y \mid X_1, ..., X_K; Z_1, ..., Z_M \thicksim \mathcal{G} (Y \mid X_1, ..., X_K; Z_1, ..., Z_M)
\label{eq:EMOS_cond2}
\end{equation}
where $Y$ is the observation, $X_1, \cdots, X_K$ are, as in the standard EMOS, the $K$ ensemble member forecasts (e.g., the wind at a specific location and look-ahead time), and $Z_1, \cdots, Z_M$ are $M$ observables identified as important to disentangle the whole forecast error.
From a more technical point of view, here the conditioning variables are categorical variables taking on one of a limited number of possible values (levels). The number of levels will be in general different for each variable and will be not fixed a priori: it will be determined in a way to minimize the CRPS (see below). All our functions of the conditioning variables will thus be categorical functions, that is,  for a given combination of  classes’ levels they will take on a constant value.\\
The form of the predictive distribution is thus exactly as in the standard EMOS apart the key fact that now $a, b_1, \cdots, b_K, c$ and $d$ in (\ref{eq:EMOS_mu}) and (\ref{eq:EMOS_sigma}) are best-fitted for each combinantion of classes' levels, via a training set, by minimizing the CRPS. Dependences on $\bm{Z}$ thus arise and will be denoted as $a(\bm{Z}), b_1(\bm{Z}), \cdots, b_K(\bm{Z}), c(\bm{Z})$, and $d(\bm{Z})$.\\
How many conditioning variables one can consider depends on the length of the training period: larger training periods, bigger values of $M$. The following set of conditioning variables has been considered in our study: the 10-m wind direction, the boundary-layer height, the hour of the day (hourly samples), the ratio between the surface wind gust and the 10-m wind speed, the ratio between 10-m and 100-m wind speed, the variance of the 10-m wind speed of the 4 grid points around the station, and the surface wind gust (all divided in classes). All of them refer to a specific location and look-ahead time.
It is worth noting that the 10-meter wind field appears
  in the conditional variables. This fact accounts in a simple way
  for nonlinearities and justifies the name of `nonlinear nonhomogeneous regression' for our strategy.\\
    It is worth remarking that our EMOS$_{+}$ strategy differs from a variant of the standard EMOS strategy described in Sec.\ \ref{The basic EMOS} where
  our conditional variables are inserted into (\ref{eq:EMOS_mu}) as $M$ additional predictors, thus changing the mean of the predictive distribution in 
  $\mu = a+b_1 X_1 +\cdots + b_K X_K + \cdots b_{K+M} Z_M $. Unlike our strategy, such a variant remains linear (i.e.\ the EMOS coefficients $ a, b_1, \cdots, b_K, \cdots , b_{K+M}$  are constant) thus being expected to possess a lower level of accuracy. We will further discuss this issue in Sec.\ \ref{Calibrations of EPS forecasts} by a direct comparison of the two different strategies.\\ Coming back to our strategy,
  
  on the basis of a cross validation carried out on the training set (year 2018) we have identified, for each station, the most effective M=3 conditioning variables, i.e.\ those for which the minimum CRPS is obtained. A larger number could be considered for larger training sets than the one considered here.
\begin{figure}[h!]
\includegraphics[width=\textwidth]{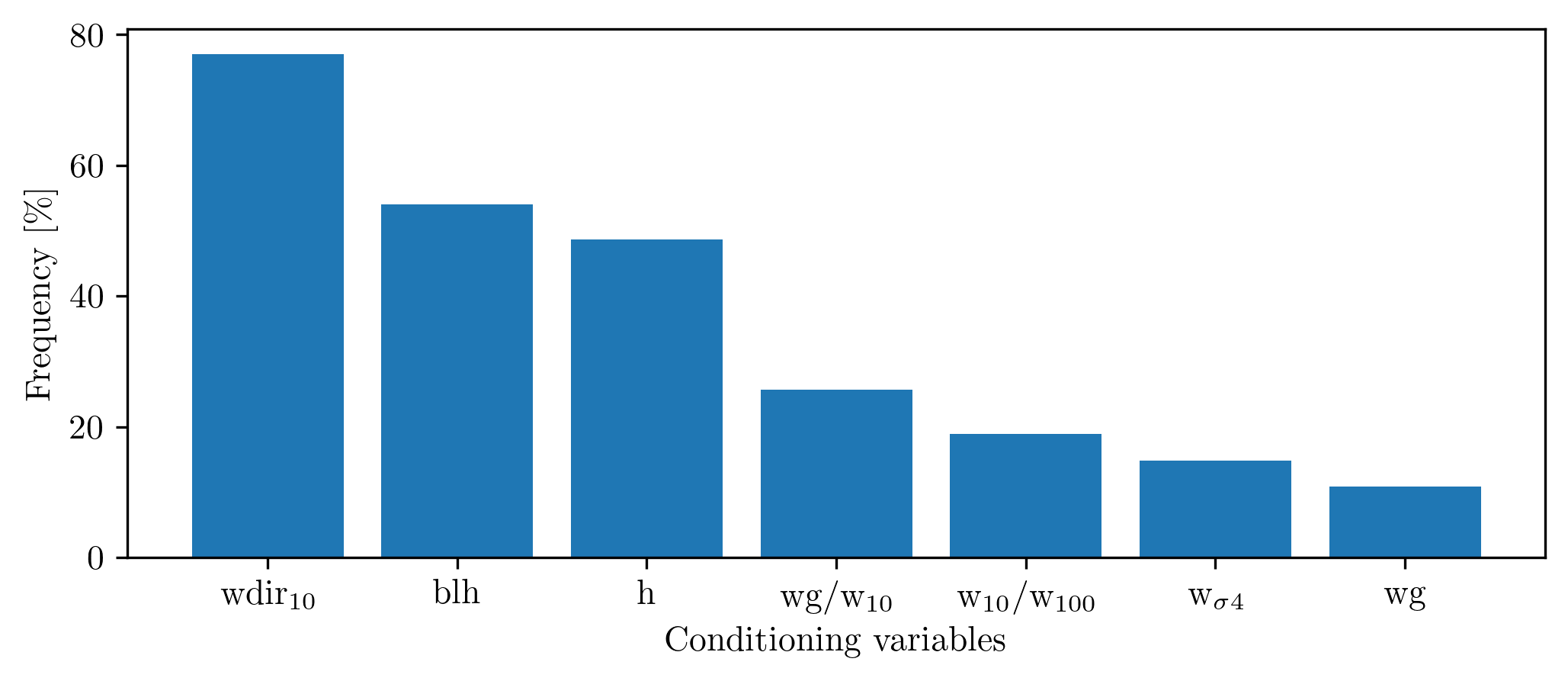}
\caption{Shown is the frequency of occurrence of the most effective  conditioning variables (i.e.\ those occurring with a frequency greater than 5$\%$). The variables shown are: $wdir_{10}$, the 10-m wind direction; $blh$, the boundary layer height; $h$, the hour of the day; $wg/w_{10}$, the ratio between the surface wind gust and the 10-m wind speed; $w_{10}/w_{100}$, the ratio between 10-m and 100-m wind speed; $w_{\sigma 4}$, the variance of the 10-m wind speed of the 4 grid points around the station;  $wg$, the surface wind gust.}
\label{fig:condizionamenti_selezionati}
\end{figure}
\noindent
Figure \ref{fig:condizionamenti_selezionati} shows the frequency with which each conditioning variable is selected. A given percentage means the number of stations where a certain conditioning variable has been selected divided by the number of the considered SYNOP stations. From the figure it appears that the most important conditioning variable is the wind direction, as expected, which is selected in most of the stations. This may be due to the complex orography characterizing the Italian territory, a fact that causes direction-dependent wind-speed-forecast errors when low-resolution prediction models are exploited.\\
The boundary-layer height and the hour of the day are also frequently selected as conditioning variables. This seems to indicate the important role of atmospheric stability to disentangle the structure of the model error.\\
The cross validation we mentioned above to select the most effective three conditioning variables, also serves to identify the optimal number of classes used to pinpoint the conditioning variables.
\begin{figure}[h!]
\includegraphics[width=\textwidth]{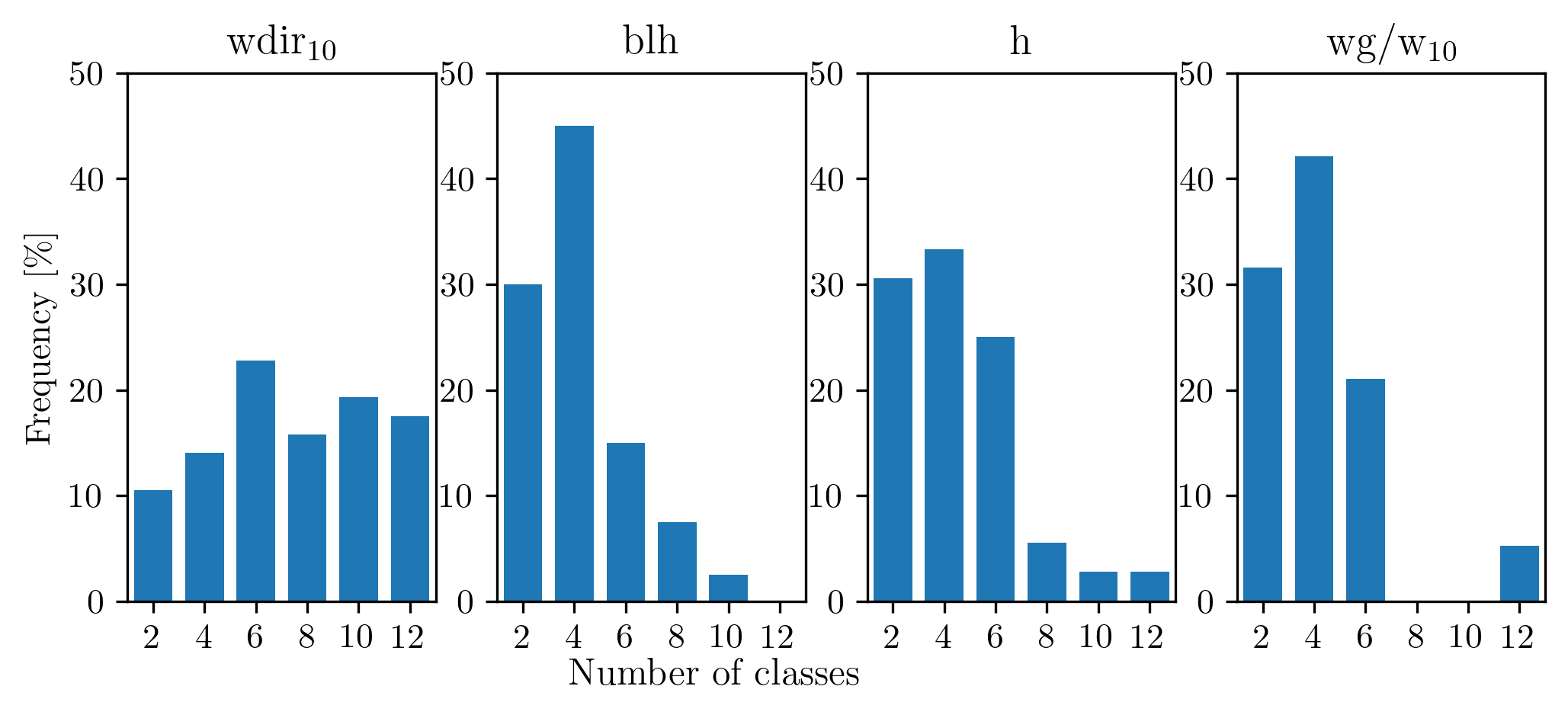}
\caption{The frequency distribution of the number of partition classes used for the
4 top conditioning variables. The 4 top variables are: $wdir_{10}$ (10-m wind direction), $blh$ (boundary-layer height), $h$ (hour of the day), and  $wg/w_{10}$ (ratio between the surface wind gust and the 10 m wind speed).}
\label{fig:numero_partizioni_condizionamenti}
\end{figure}
\noindent
Figure \ref{fig:numero_partizioni_condizionamenti} shows the frequency distributions of the number of classes used for the sampling of the selected conditioning variables.  Only the distributions for the 4 top conditioning variables are considered.
 For each variable, the number of considered classes ranges from 2 to  12. From the figure it can be seen that the wind direction is not only the most used conditioning variable but is also the one often requiring the densest sampling. This fact further confirms the possible role of sub-grid orographic variations as possible cause of direction-dependent errors on the wind-speed forecast.

\subsubsection{The EMOS$_4$ strategy}
\label{The EMOS_4 strategy}

The EMOS$_+$ strategy is further generalized to account for possible forecast errors due to a mismatch between the station position and the selected model grid point. The issue is potentially important especially for model strategies, as the EPS, the spatial resolution of which are relatively coarse.
Accordingly, many subgrid orographic features and details of the coastline  are lost by the EPS and one needs to insert them via a suitable calibration. But what is the model gridpoint best representing the ground station?  The one closest to the station or another among the four closer to it? Answering this question is not obvious a priori, especially in situations of complex orography and/or complex coastlines  where the model gridpoint closest to the ground station could be swept right out to sea.  Our strategy deals with this issue and is able to identify the most representative model gridpoint.\\
In more details, the predictive probability is still conditional on $ \mathbf{X} = (X_1, \cdots, X_K)$ and $\mathbf{Z} = (Z_1, \cdots, Z_M)$ as for the EMOS$_+$ strategy but now its mean value is:

\begin{equation}
  \mu   = a(\mathbf{Z}) + \sum_{i=1,j=1}^{K,4} b_{ij}(\mathbf{Z}) X_i^j 
\label{eq:EMOS_mu_cond}
\end{equation}
\begin{equation}
  \sigma   = c(\mathbf{Z}) + \sum_{j=1}^{4} d_{j}(\mathbf{Z}) S^{2 \: j} 
\label{eq:EMOS_sigma_cond}
\end{equation}

\noindent
where $j$, the upper index, spans over the 4 model grid points around the station. Eqs.\ (\ref{eq:EMOS_mu_cond}) and  (\ref{eq:EMOS_sigma_cond}) are, in general, nonlinear relationships between predictands and predictors: the 10-meter wind speed also enters as conditioning variable.\\
For the sake of clarity, $X_i^j$ now denotes the i-th ensemble member forecast on the j-th model grid point (among the four identified around the stations) while $S^{2 \: j}$ denotes the variance of the j-th model grid point.\\
The predictive distribution here is conditional on the values of $Z_1, \cdots, Z_M$ averaged on both the 4 nodes around the station and the $K$ ensemble member forecasts.
We dub the strategy as EMOS$_{4}$.\\
A further strategy we will exploit is to apply, downstream to the EMOS$_+$ carried out on each of the 4 nodes around the station, a further EMOS$_+$ to determine the optimal mix of outcome on the 4 nodes.\\
In plain word, this is done with a predictive gamma function with mean and variance given by:

\begin{equation}
  \mu = a(\mathbf{Z}) + \sum_{i=1,j=1}^{K,4} b_{ij}(\mathbf{Z})\widetilde{X}_i^j
\label{eq:EMOS_mu_cond2}
\end{equation}
\begin{equation}
  \sigma = c(\mathbf{Z}) + \sum_{j=1}^{4} 
d_{j}(\mathbf{Z})\widetilde{S}^{2 \: j}
\label{eq:EMOS_sigma_cond2}
\end{equation}

where $\widetilde{X}_i^j$ is the EMOS$_+$ calibrated i-th ensemble-member forecast on the j-th model grid point and $\widetilde{S}^{2 \: j}$ is the ensemble spread on the j-th model grid point and as for the EMOS$_4$ strategy the predictive distribution is conditional on the values of $Z_1, \cdots, Z_M$ averaged on both the 4 nodes around the station and the $K$ ensemble member forecasts.
The discrete calibrated samples $\widetilde{X}_i^j$ can be easily obtained via the so-called  Empirical Copula Coupling (ECC-Q) described by \cite{schefzik2013uncertainty} once the parameters in the predictive EMOS distribution are determined. We dub this approach as EMOS$_{+4}$.

\subsubsection{The EMOS$_r$ strategy}
\label{The EMOS_r strategy}

Our choice to make the predictive distribution conditional on $\mathbf{Z}$ other than on $\mathbf{X}$ implies the need of having at disposal large training set. For this reason we found reasonable to consider at least a one-year long training set. This is somehow in contrast with a sort of solid protocol to consider rolling windows of 40 days ending at the day of the forecast emission \citep{gneiting2005calibrated} as a reliable optimal training set.\\
Our idea is thus to merge the advantages of the two different strategies. The rolling window naturally allows one to synchronize to the current climate trend, a fact that seems less effective for a ``static" training set. The static training set, on the other hand, has the advantage to permit heavy conditioning on the predictive distribution.\\
A possible way to maintain both strengths is to apply downstream to our EMOS$_{+4}$ strategy a further EMOS 
with 40-day rolling training set (without conditioning on $\mathbf{Z}$).
In more details, one has to apply the EMOS$_0$  strategy described in Sec.~3.1 where now the raw EPS members, $X_1, \cdots ,  X_K$, must be replaced by their  EMOS$_{+4}$-calibrated counterpart. As in Sec.\ \ref{The EMOS_4 strategy},
  the discrete calibrated ensemble samples have been obtained via Empirical Copula Coupling once the parameters in the predictive EMOS distribution are determined. \\
The final result is called here EMOS$_{+4r}$, where $r$ stands for ``rolling''.

\section{Statistical indices to assess the quality of the calibration}
\label{Statistical indexes to assess the quality of the calibration}

To evaluate the effectiveness of the different EMOS strategies, we compare the resulting calibrated forecasts against reference forecasts consisting either of one of the EMOS approaches proposed here or of simpler predictions based on persistence or climatology.\\
To make the comparison as quantitative as possible, the Skill Score (SS) index \citep{wilks2011statistical} will be used. The skill score indicates how better the calibrated forecast is with respect to a reference forecast. Lower bounds vary depending both on the score used  to compute the skill and on the reference forecast used. Negative values of SS means that the calibrated forecast is of lower quality than the reference one. Upper bounds are always 1. $SS=0$ indicates no improvement over the reference forecast, while 1 indicates perfect performance. In quantitative terms, the skill score is defined as:

\begin{equation} 
SS = \frac{A - A_{ref}}{A_{opt} - A_{ref}}
\label{eq:Skill_Score}
\end{equation}
where $A$ is the error index value of the calibrated forecast, $A_{ref}$ is the error index value of the reference forecast, and $A_{opt}$ is the optimal index value.\\
The error indices used here are the normalized mean absolute error (NMAE), the correlation coefficient and the so-called reliability index, $\Delta$, proposed by \cite{delle2006probabilistic}.\\
The normalized mean absolute error (NMAE), often preferred to the more classic NRMSE \citep{willmott2005advantages}, is an arithmetic average of the absolute values of the differences between each pair members \citep{wilks2011statistical} normalized with the mean of the observations. 
Considering $Y_n$ the n-th observation and $X_n$ the corresponding n-th forecast (here corresponding to the mean of the 50 EPS ensemble), the NMAE is defined as:

\begin{equation}
NMAE = \frac{\sum_{n=1}^{N}\left|X_n - Y_n\right|}{\sum_{n=1}^{N}Y_n}
\label{eq:NMAE}
\end{equation}
where $N$ is the number of observation-forecast pairs in a given test set.\\
The correlation coefficient, $\cal{C}$, is a measure of linear dependence between two variables \citep{wilks2011statistical} and takes values between -1 and 1, where 1 represents the maximum correlation, -1 the maximum anti-correlation. In plain formula \citep{lee1988thirteen},

\begin{equation} 
{\cal{C}} = \frac{\sum_{n=1}^{N}(X_n-\overline{X})(Y_n-\overline{Y})}{N\sigma_X \sigma_Y}
\label{eq:Pearson}
\end{equation}
with:
\begin{equation}
\sigma_X = \sqrt{\frac{\sum_{n=1}^{N}(X_n-\overline{X})^2}{N}}
\end{equation}
\begin{equation}
\sigma_Y = \sqrt{\frac{\sum_{n=1}^{N}(Y_n-\overline{Y})^2}{N}}
\end{equation}
where $\overline{X}$ and $\overline{Y}$ are the mean values of $X$ and $Y$.\\
As stated by \cite{gneiting2007probabilistic}, the aim of the probabilistic forecast is to maximize the sharpness of the predictive distribution subject to calibration. Verification rank (VR) histograms are a graphical tool proposed by \cite{anderson1996method} and \cite{hamill1997verification} to assess the calibration of ensemble forecasts. When the ranks of the observations are pooled within the ordered ensemble forecasts, VR histograms show the distribution of the ranks.
The observations and ensemble predictions should be interchangeable in a calibrated ensemble, resulting in a uniform VR histogram. The probability integral transform (PIT) histogram is the continuous analogue of the VR histogram \citep{dawid1984present, diebold1997evaluating, gneiting2007probabilistic}. The value of the predictive cumulative distribution function at the verifying observation defines the PIT value.
The empirical cumulative distribution function of the PIT values should converge to the  uniform distribution for calibrated forecasts.\\
\cite{delle2006probabilistic} propose the reliability index $\Delta$ to quantify the deviation of VR histograms from uniformity. We will use the following definition of $\Delta$ to quantify the deviation from uniformity in the PIT histograms:

\begin{equation} 
\Delta = \sum_{i=1}^{m}\left|f_i - \frac{1}{m}\right|
\label{eq:Delta}
\end{equation}
where $m$ is the number of classes in the histogram, each with a relative frequency of $1/m$, and $f_i$ is the observed relative frequency in class $i$.\\
This index varies between 0 and $+\infty$: the closer to zero the better the calibration.


\section{Calibration for the wind speed}
\label{`Local' calibration for the wind speed}

\subsection{Forecasts using observations: persistence and climatology}
\label{Forecasts using observations: persistence and climatology}
Persistence and climatology are forecasts based solely on the observed data that each station has access to. They are especially useful when predictive models are not available; they will be used here as baseline forecasts to evaluate the added value of different calibration strategies.\\
Persistence-based forecasts can be defined in two different ways. The first one is based on the diurnal cycle: the forecast at a given look-ahead time $t$ consists of the observations at time $t-24$ hours ($t-48$ hours when the look-ahead time is in the 24-48 hour forecast interval). The second one is based on the most basic definition of persistence, in which the forecast n-hour ahead with respect to a conventional starting date (here 00 UTC) is simply given by the observation at 00 UTC.\\
The EMOS$_+$ strategy was used to calibrate these two raw persistence-based forecasts on the basis of a training set the length of which varies between 2 and 5 years, depending on the data availability. Because our forecast based on persistence is not an ensemble of forecasts, but just one forecast, the variance of the predictive distribution has been calculated only in terms of the parameter $c$ in Eq.~(\ref{eq:EMOS_sigma}), leaving the variance to be a function of the conditionals.\\
\begin{figure}[h!]
\includegraphics[width=\textwidth]{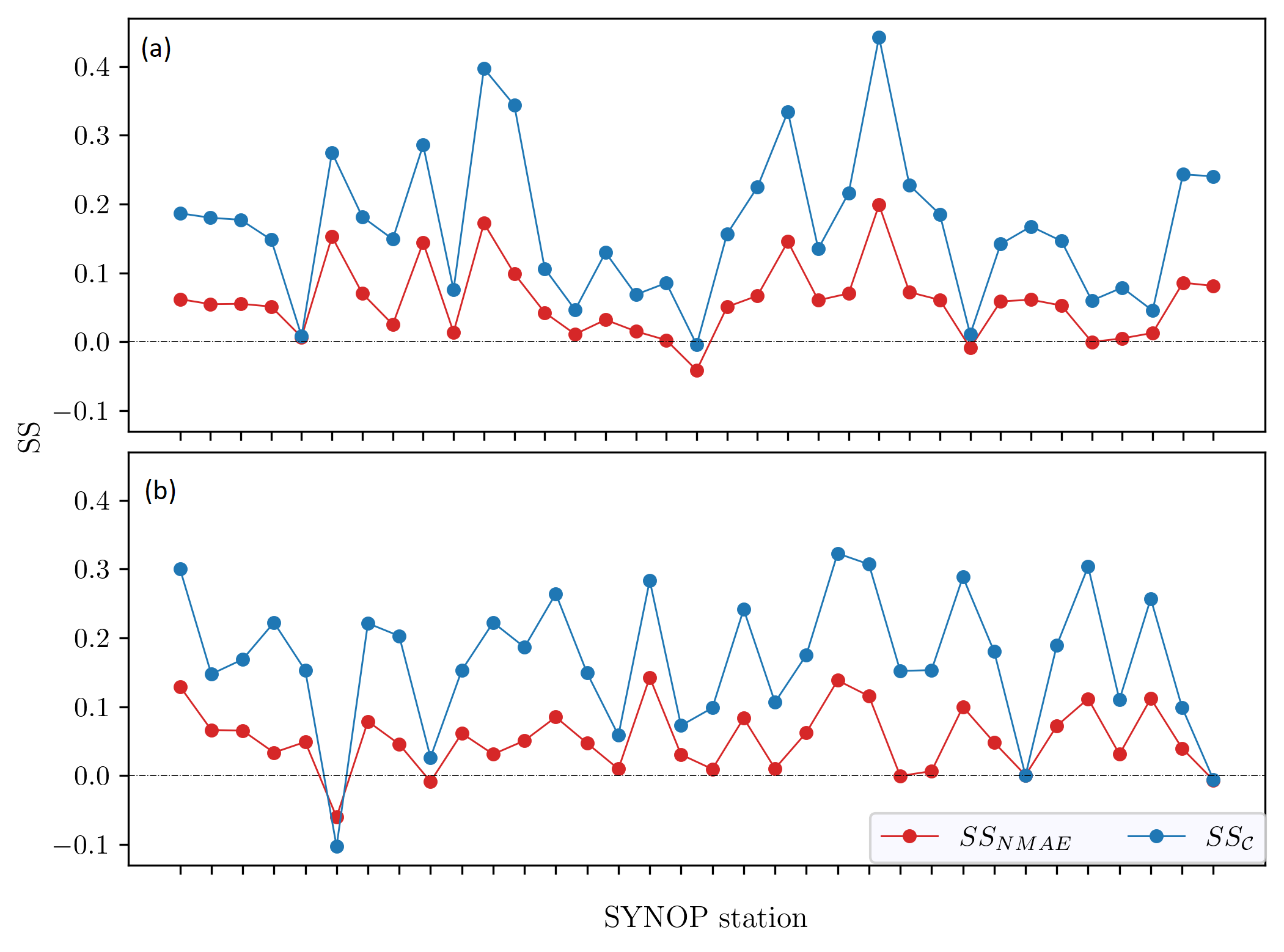}
\caption{Skill score of NMAE (red lines) and correlation coefficient (blue lines), $\cal{C}$, relative to a 24-hour look ahead time forecast based on persistence for different SYNOP stations (panel (a): stations from 1 to 35; panel (b): stations from 36 to 69). Persistence is defined here for all look-ahead times as the observed value at time 00 UTC. This simple forecast is successively calibrated in terms of our EMOS$_+$. The reference forecast to compute the skill scores is the calibrated (via EMOS$_+$) forecast based on the persistence where for any look-ahead time the forecast is the observation 24 hours before. Note how the persistence based on the 00 UTC observations performs better than persistence based on observations 24 hours before the look-ahead time. This is true both in terms of the NMAE (mean SS$_{NMAE}$ = 6$\%$) and in terms of correlation coefficient (mean SS$_{\cal{C}}$ = 17$\%$). For the 24-48 hour interval forecast the results are opposite (not shown): the mean NMAE ($\cal{C}$) skill score of the daily-cycle-based persistence with respect to the 00 UTC-based persistence is 3$\%$ (4$\%$).}
\label{fig:persistenze_calibrate}
\end{figure}
\noindent
For all 69 available SYNOP stations, Figure\ \ref{fig:persistenze_calibrate} shows a comparison of the two forecasts based on persistence after calibration with the EMOS$_+$ strategy in which persistence based on the diurnal cycle is taken as the reference forecast. Both NMAE and correlation coefficient demonstrate that the persistence that uses the most recent available observation (here 00 UTC) leads to a better prediction; the NMAE shows an average skill score of 6$\%$ 
while for the correlation coefficient the average skill score is about 17$\%$.
The conclusion is opposite in the 24-48 hour forecast interval, where the daily-cycle-based persistence works better.\\
Note that, having forecasts for each hour, here the scores are obtained as an average over the whole 24-hour period, one corresponding to the interval 0-24 h and the other to the interval 24-48 h.
\begin{figure}[h!]
\includegraphics[width=\textwidth]{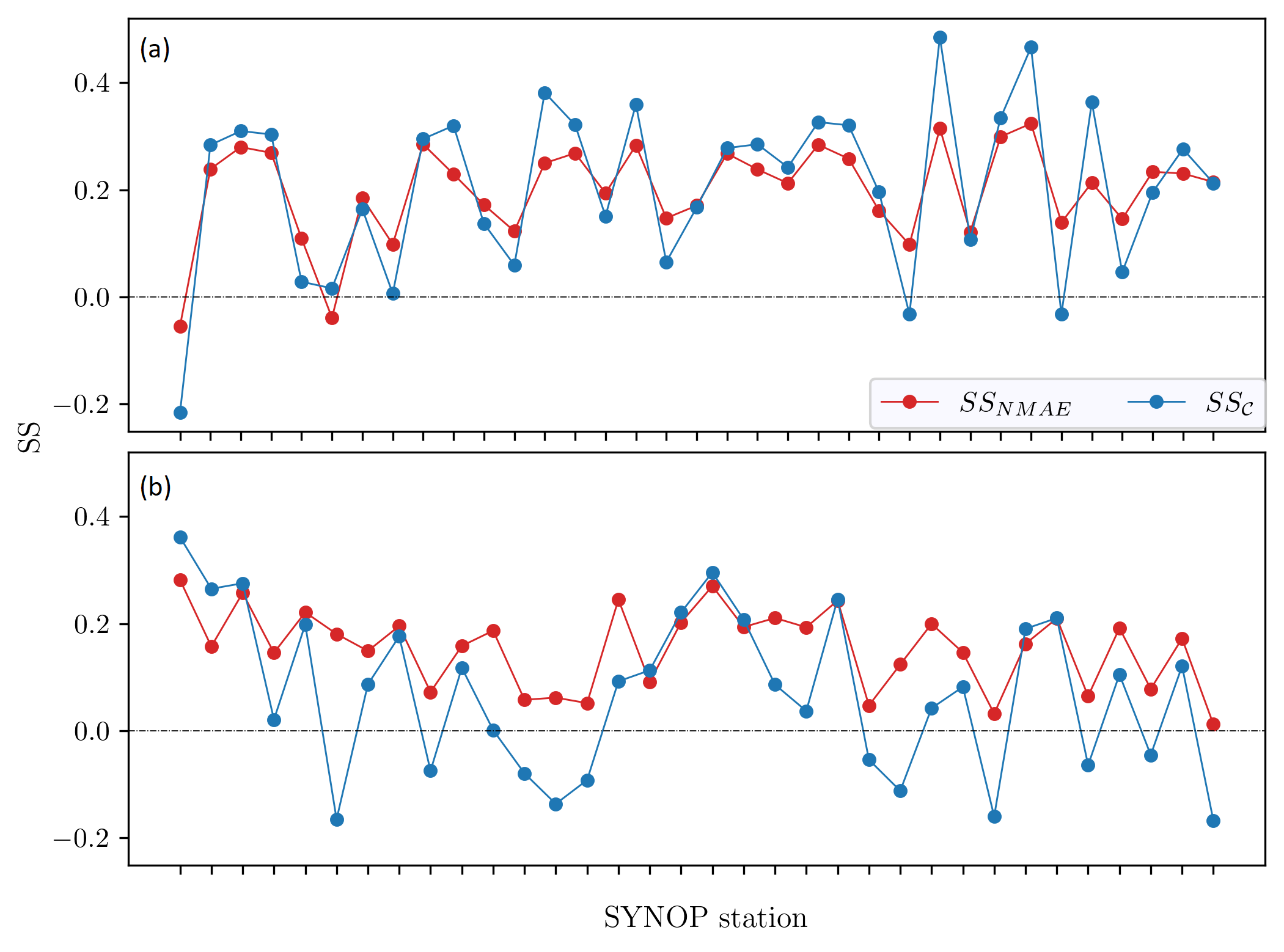}
\caption{Skill score of NMAE and correlation coefficient relative to a 24-hour look ahead time forecast based on persistence for different SYNOP stations (panel (a): stations from 1 to 35; panel (b): stations from 36 to 69). Here, persistence is defined for all look-ahead times as the observed value at time 00 UTC. This simple forecast is successively calibrated in terms of our EMOS$_+$. The reference forecast to compute the skill scores is the raw persistence defined for all look-ahead times as the observed value at time 00 UTC. Note how the calibrated persistence-based forecast performs better than the raw persistence-based forecast. This is true both in terms of NMAE (mean SS$_{NMAE}$ = 18$\%$) and in terms of the correlation coefficient (mean SS$_{\cal{C}}$ = 14$\%$). For the 24-48 hour interval forecast (not shown), the mean NMAE skill score is 23$\%$ and the mean correlation coefficient is 15$\%$.}
\label{fig:persistenza00utc_prima_dopo_calibrazione}
\end{figure}
\noindent
To evaluate the contribution made by the EMOS$_+$ calibration, Figure\ \ref{fig:persistenza00utc_prima_dopo_calibrazione} shows a comparison of the
persistence that uses the observations at 00 UTC
(0-24 hour forecast) before and after calibration, using the raw persistence as the reference forecast. With an overall improvement of nearly 18$\%$, the NMAE is the index showing the largest improvement.\\
Let us now compare the
persistence that uses the observations at 00 UTC
 against climatology to identify the best strategy among the two. Climatology has been calculated by constructing a typical day representative of each month of the year having hourly variability on the basis of a training set of 2/4 years.

\begin{figure}[h!]
\includegraphics[width=\textwidth]{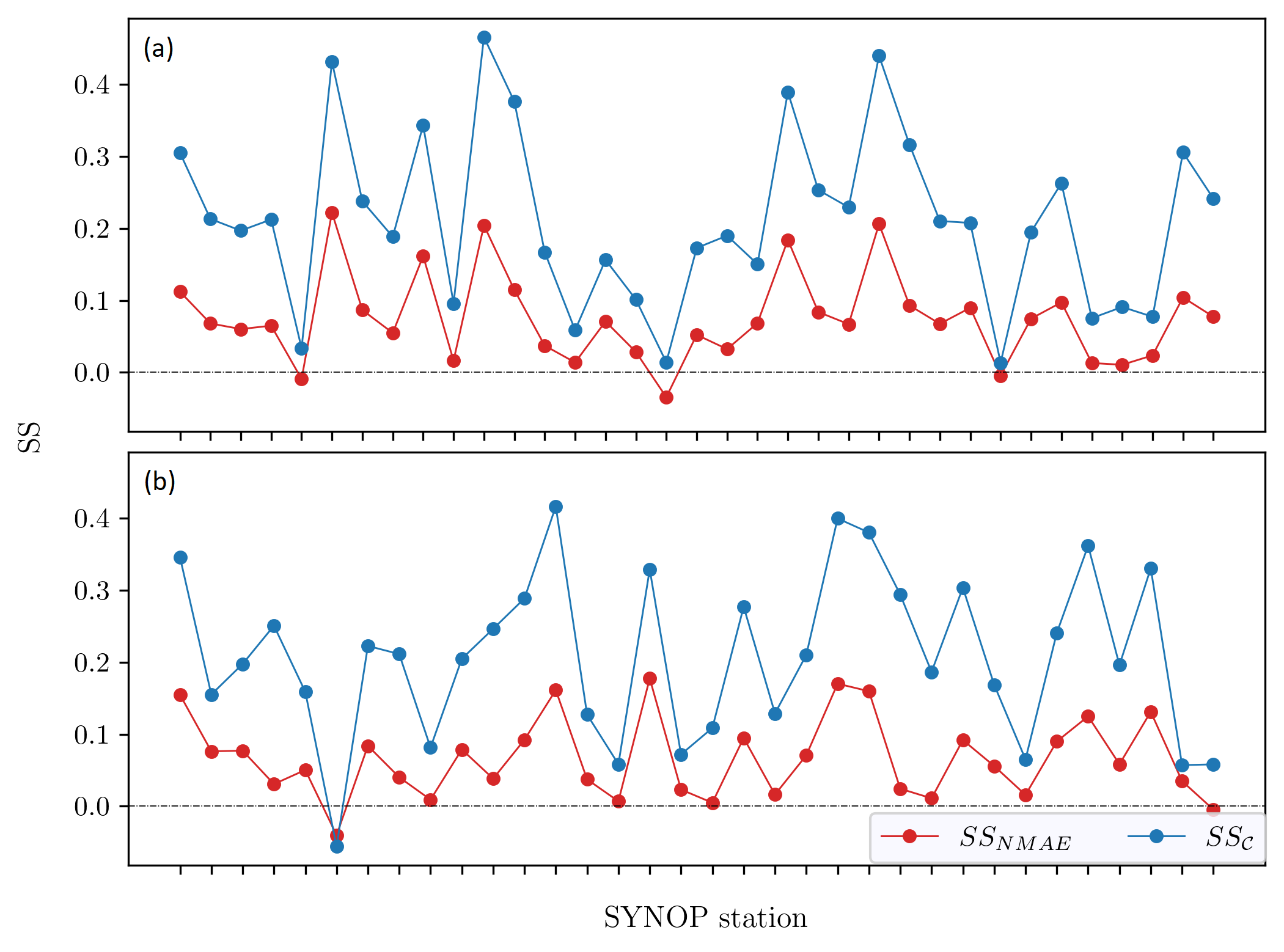}
\caption{Skill score of NMAE and correlation coefficient relative to a 24-hour look ahead time forecast based on persistence for different SYNOP stations (panel (a): stations from 1 to 35; panel (b): stations from 36 to 69). Here, persistence is defined for all look-ahead times as the observed value at time 00 UTC. This simple forecast is successively calibrated in terms of our EMOS$_+$. The reference forecast to compute the skill scores is the forecast based on climatology. Note how the calibrated persistence-based forecast performs better than climatology. This is true both in terms of the NMAE (mean SS$_{NMAE}$ = 7$\%$) and in terms of the correlation coefficient (mean SS$_{\cal{C}}$ = 21$\%$). For the 24-48 hour forecast interval (not shown), climatology outperforms persistence having a mean skill score of 3$\%$ (for NMAE) and 6$\%$ (for $\cal{C}$).}
\label{fig:persistenza00utc_Epl_vs_climatologia}
\end{figure}
\noindent
Figure\ \ref{fig:persistenza00utc_Epl_vs_climatologia} shows the comparison between the two forecasts using climatology as the reference forecast. Except for a very few cases, persistence is better in all stations, as expected.\\
The mean skill score is about 7$\%$ for the NMAE and about 21$\%$ for the correlation coefficient. Climatology overcomes persistence in the 24-48 hour forecast interval (not shown). Accordingly, the skill score of climatology using the best forecast based on persistence as reference has a mean NMAE of about 3$\%$ and a mean correlation coefficient of about 6$\%$.\\
As we have already discussed, Figures \ref{fig:persistenze_calibrate}, \ref{fig:persistenza00utc_prima_dopo_calibrazione}, \ref{fig:persistenza00utc_Epl_vs_climatologia} show forecast improvements in terms of the NMAE index and of the correlation coefficient. However, these indices only assess the accuracy of the mean of the ensemble. We need to resort to the PIT histogram to quantify the ensemble forecast accuracy as a whole.

\begin{figure}[h!]
\includegraphics[width=\textwidth]{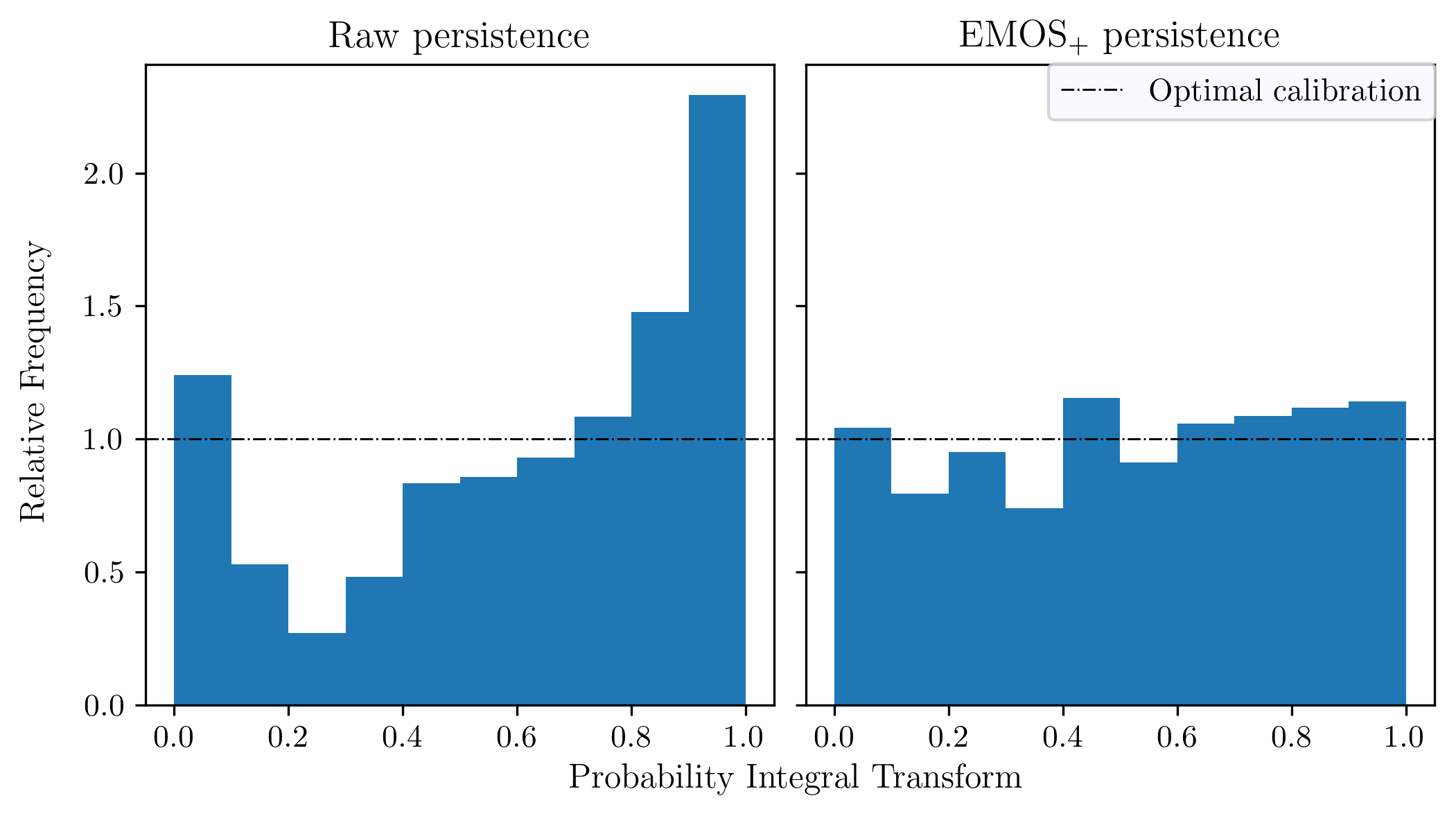}
\caption{PIT histogram of the forecast based on persistence (with observation at 00 UTC) before (left panel) and after (right panel) the EMOS$_+$-based calibration. The histogram refers to the SYNOP station having a value of the index $\Delta$ corresponding to the median. The average value of the index $\Delta$ of all stations for the raw persistence is 0.50 while for the calibrated persistence is 0.14 signaling a clear improvement due to the EMOS$_+$ calibration. For the 24-48 hour forecast interval (not shown), the mean $\Delta$ value passes from 0.53 (raw) to 0.17 (calibrated).}
\label{fig:PIT_persistenza00utc_prima_dopo_calibrazione}
\end{figure}

\noindent
As an example, Figure\ \ref{fig:PIT_persistenza00utc_prima_dopo_calibrazione} shows the PIT histogram of the 00 UTC-based persistence for the SYNOP station having a value of the index $\Delta$ corresponding to the median after the EMOS$_+$ calibration (right panel). The left panel refers to the same station before the forecast calibration (raw prediction) where the variance has been obtained comparing the forecast with the observed data in the same training set of 2/4 years used for its calibration.
The numbers of bins used to show the PIT diagram is 10 in all reported cases.
The PIT of the raw persistence confirms that, prior to calibration, the persistence-based forecast underestimates the mean wind speed because most observations fall in the ninth percentile of the forecast distributions. Moreover, a clear underdispersion for the ensemble variance is also evident: observations, indeed, populate more the extreme percentile values of the forecast distributions. EMOS$_+$ corrects both problems. Including all the other stations, the average value of the index $\Delta$ passes from 0.5 (uncalibrated forecast) to 0.14 (EMOS$_+$ calibration).

\subsection{Calibrations of EPS forecasts}
\label{Calibrations of EPS forecasts}

We will now analyze the results starting from the 24-h raw forecast to arrive at the best calibration strategy, i.e. our EMOS$_{+4r}$. The accuracy of the average wind forecast from the EPS will be assessed also in this case via the NMAE and correlation coefficient. The results will be presented for all 69 SYNOP stations, as before.\\
\begin{figure}[h!]
\includegraphics[width=\textwidth]{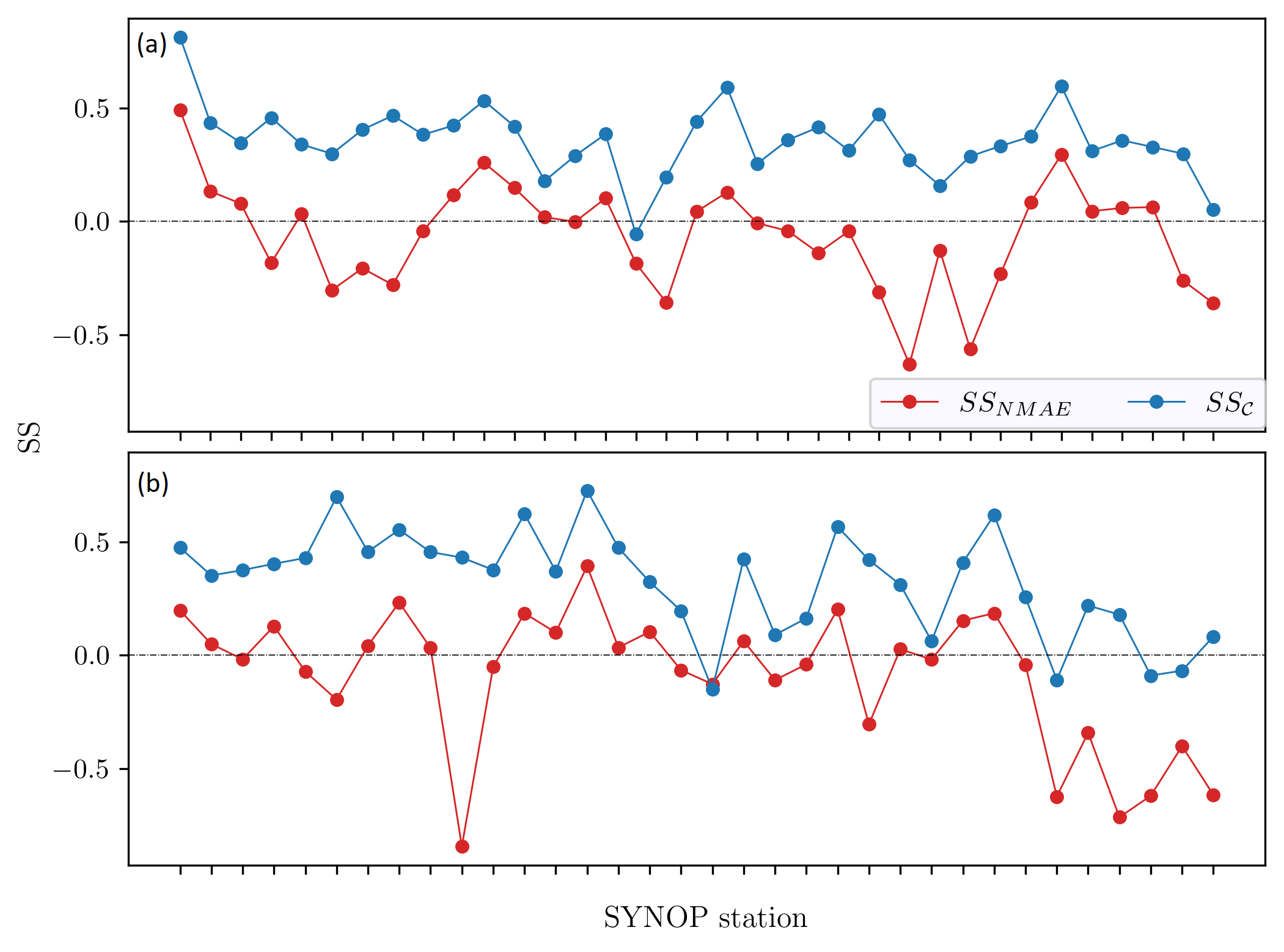}
\caption{Skill score of the 10 m raw EPS-based wind forecast using the calibrated best persistence as a reference for different SYNOP stations (panel (a): stations from 1 to 35; panel (b): stations from 36 to 69). The graph shows a different result for the two indices: for the NMAE the best prediction is the one based on the persistence since on average SS$_{NMAE}$ is negative (the mean SS$_{NMAE}$ of the
persistence-based forecast that uses the observations at 00 UTC
with respect to the 10 m raw EPS-based forecast is 4$\%$). The correlation coefficient is better for the raw EPS forecast with an average SS of 34$\%$. This means that, calibrated persistence provides a forecast more accurate than the raw EPS forecast. Also for the 24-48 hour forecast interval (not shown), the mean NMAE skill score is negative (the mean SS$_{NMAE}$ of the
persistence-based forecast that uses the observations at 00 UTC
with respect to the 10 m raw EPS-based forecast is -5$\%$) while SS$_{\cal{C}}$ is 49$\%$. The same conclusions drawn for the 0-24 hour forecast interval thus hold also for the 24-48 hour forecast interval.}
\label{fig:persistenza00Epl_vs_prev10m}
\end{figure}
\noindent
Let us start from a comparison between the raw EPS forecast and the
persistence-based forecast that uses the observations at 00 UTC. Figure\ \ref{fig:persistenza00Epl_vs_prev10m} depicts the NMAE and correlation coefficient skill scores for the raw EPS-based forecast, with the
persistence-based forecast that uses the observations at 00 UTC
 serving as reference forecast. The two skill scores lead to opposite conclusions. Indeed, the raw EPS forecast has a higher correlation coefficient than persistence-based forecast (the mean value of the $\cal{C}$ skill score for 0-24 hour forecast interval is 34$\%$ and for 24-48 hour forecast interval is 49$\%$) but has a lower NMAE. The same result as for the NMAE was found for $\Delta$ (not shown) with a negative average skill score. Accordingly, computing the skill score of the persistence-based forecast using the raw EPS-based forecast as reference, the average skill score of $\Delta$ for the 0-24 hour forecast interval is 86$\%$ while for the 24-48 hour forecast interval is 86$\%$. For the mean NMAE skill score we have 4$\%$ (0-24 hour forecast interval) and -5$\%$ (24-48 hour forecast interval).\\
Despite the larger NMAE of the raw EPS forecast, the fact that it has a larger correlation coefficient suggests the possibility to obtain relevant improvements even in terms of simple calibration strategies as the EMOS$_0$. We will address this issue in the following.\\
When addressing the issue of the calibration of the EPS-based forecast, one has firstly to identify the model observables best representing the observed wind. The EPS indeed provides wind forecasts at different elevations above the ground (here, we only retain 10 m and 100 m) together with the 10 m wind gust accounting for short-time wind temporal fluctuations. Because of the coarse spatial resolution of the EPS, the selection of the most representative observable is not obvious a priori.\\
\begin{figure}[h!]
\includegraphics[width=\textwidth]{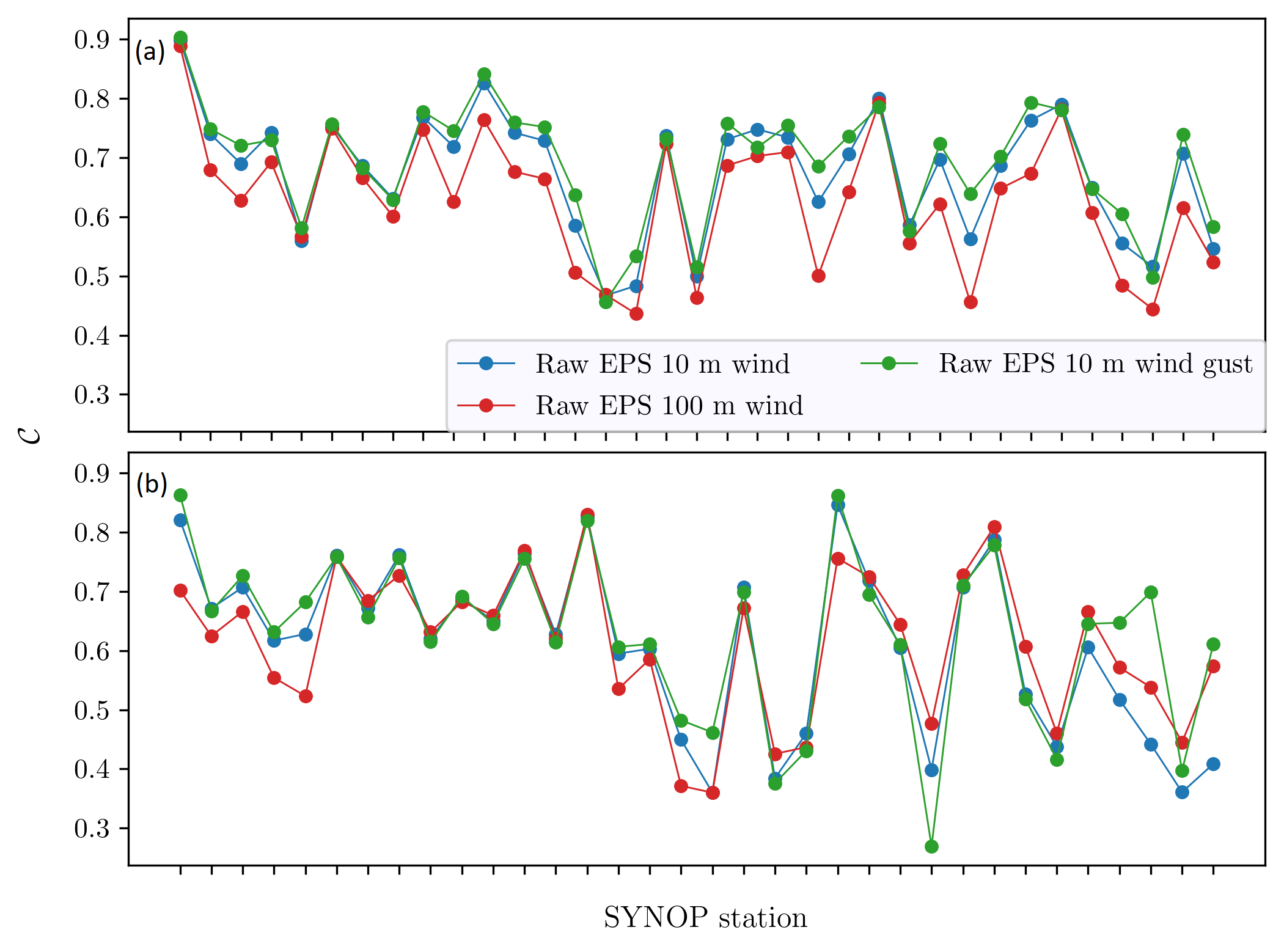}
\caption{Shown is the correlation coefficient, $\cal{C}$, for the following three raw forecasts: the 10 m wind forecast, the 100 m wind forecast, and the 10 m wind gust forecast for different SYNOP stations (panel (a): stations from 1 to 35; panel (b): stations from 36 to 69). There is no option best-performing for all stations. The selection of the best option will be thus performed separately for each station on the basis of a suitable training set.}
\label{fig:prev10_prev100_i10fg_pearson}
\end{figure}
\noindent
To give an idea of this issue, we have reported in Figure \ref{fig:prev10_prev100_i10fg_pearson} three different raw EPS forecasts based on the 10 m and 100 m wind forecasts, and on the 10 m wind gust, all interpolated at the SYNOP stations. As one can see from this figure, no general conclusion can be drawn on which observable is best representing the actual wind speed. For that reason, the choice among the three possible options (10 m wind speed forecast, 100 m wind speed forecast, 10 m wind gust forecast) has been carried out separately for each station.\\
Once the most representative model variable of the observed wind speed is determined, an initial calibration exploiting the EMOS$_0$ strategy has been performed using one year of training without using other variables for conditioning.

\begin{figure}[h!]
\includegraphics[width=\textwidth]{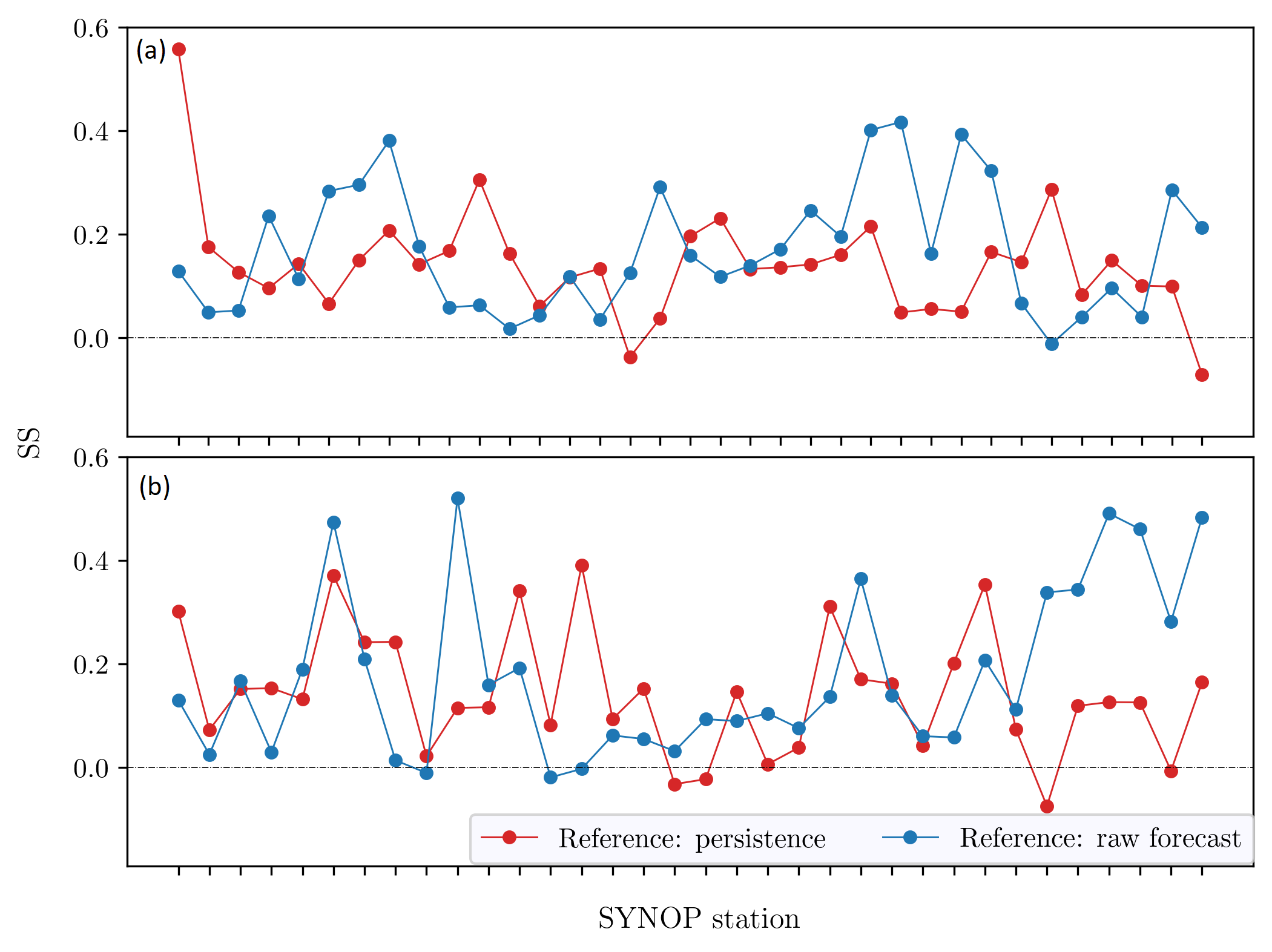}
\caption{Shown is the NMAE skill score of the EMOS$_0$ calibration using two different forecasts as reference for different SYNOP stations (panel (a): stations from 1 to 35; panel (b): stations from 36 to 69). One is the calibrated
persistence-based forecast that uses the observations at 00 UTC
 (red curve) while the other is the raw 10 m EPS wind forecast (blue curve). As expected, the calibration in terms of the EMOS$_0$ appreciably improves the EPS forecast both with respect to persistence and with respect to the raw EPS forecast. The average skill score is 14$\%$ (persistence as reference) and 17$\%$ (raw EPS as reference). For the 24-48 hour forecast interval (not shown) the average skill score is 21$\%$ (persistence as reference) and 16$\%$ (raw EPS as reference).}
\label{fig:prevE0_norm_con_pers_in_rosso_e_prev10_in_blu}
\end{figure}
\begin{figure}[h!]
\includegraphics[width=\textwidth]{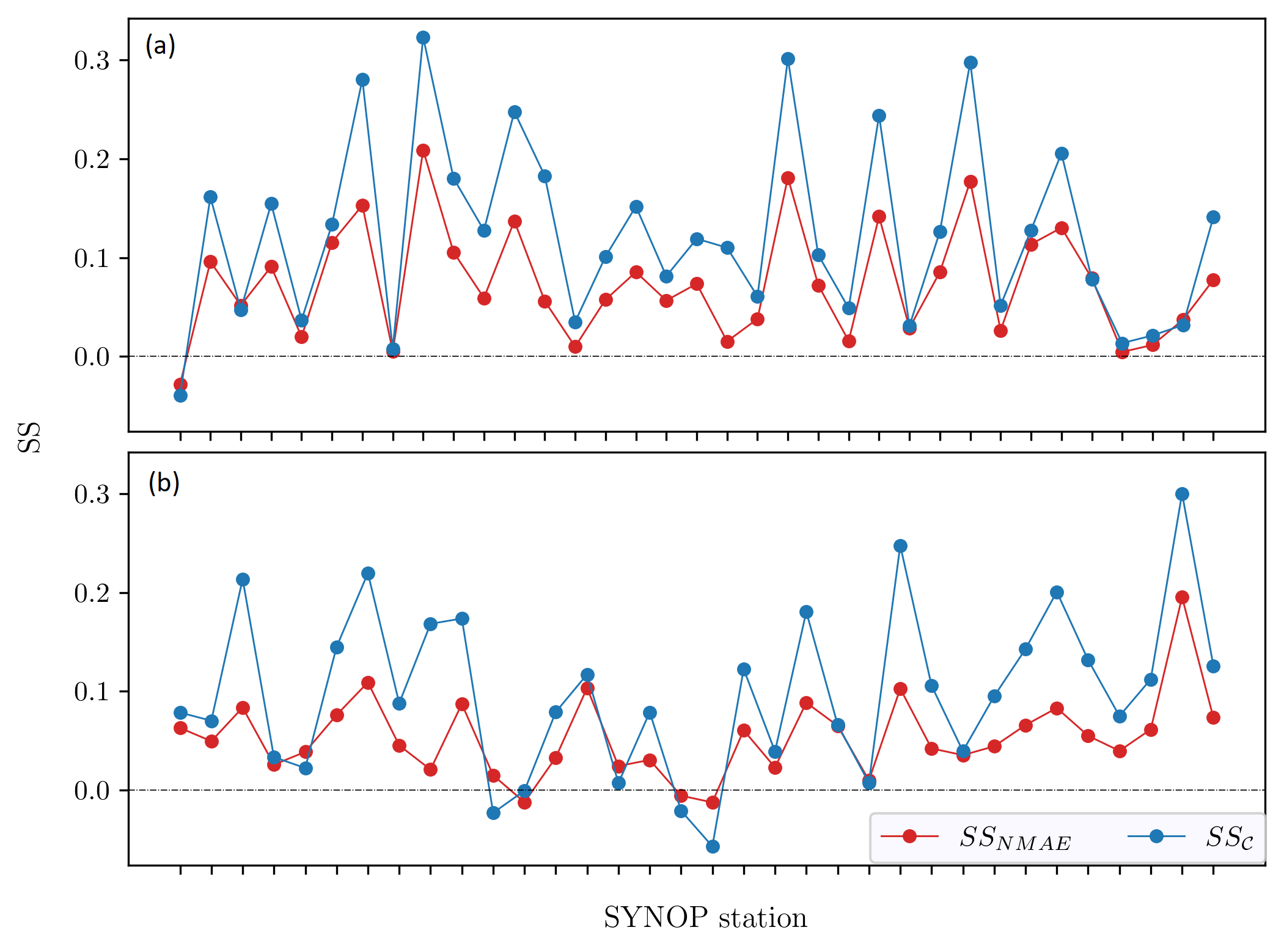}
\caption{Skill score of the forecast from our EMOS$_+$ calibration using the EMOS$_0$ variant, with our conditional variables serving as additional predictors,
  as a reference.
  Panel (a): stations from 1 to 35; panel (b): stations from 36 to 69. In almost all stations EMOS$_+$ outperforms the EMOS$_0$ variant.}
\label{fig:Epl_vs_Ereferee}
\end{figure}

\noindent
The NMAE skill score of the EPS-based forecast calibrated with the EMOS$_0$ using the best persistence and the raw wind speed EPS forecasts at 10 m as reference forecasts, respectively, is shown in Figure\ \ref{fig:prevE0_norm_con_pers_in_rosso_e_prev10_in_blu}. As expected, the EMOS$_0$-based calibration provides the best performance for the large majority of analyzed stations.\\
Let us now pass to evaluate the added value of our EMOS$_+$ calibration, thus quantifying the role of nonlinearities accounted for in our strategy. Accordingly, we will proceed in two steps. We first assess the skill of  our EMOS$_+$ calibration against the standard EMOS$_0$ involving the sole wind speeds as predictors. Secondly, the comparison will be done against the variant of the  EMOS$_0$ discussed in Sec.\ \ref{The EMOS_+ strategy} where
  all our conditional variables are inserted in the linear regression for the mean $\mu$ as additional predictors. In doing that, all variables have been normalized with their respective mean values obtained in the training set.
\begin{figure}[h!]
\includegraphics[width=\textwidth]{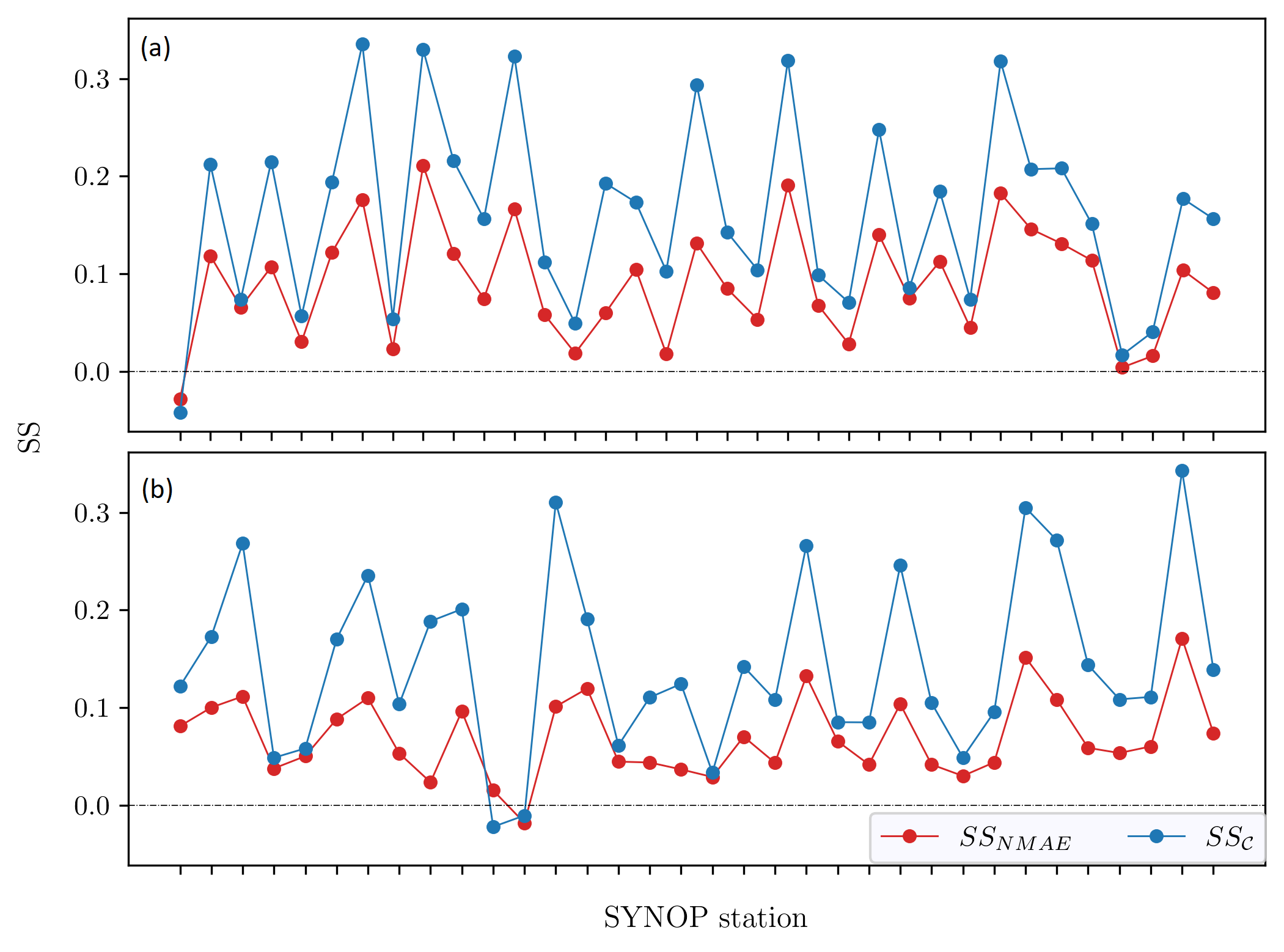}
\caption{Skill score of the forecast from our EMOS$_+$ calibration using the EMOS$_0$ calibration as the reference forecast for different SYNOP stations (panel (a): stations from 1 to 35; panel (b): stations from 36 to 69). In almost all stations there is a clear improvement, with average skill scores of 8$\%$ (NMAE) and 15$\%$ (correlation coefficient). For the 24-48 hour forecast interval (not shown) the average skill scores are 7$\%$ for the NMAE index and 13$\%$ for the correlation coefficient.}
\label{fig:Epl_vs_E0}
\end{figure}
\noindent
Figure\ \ref{fig:Epl_vs_E0} shows the NMAE and the correlation coefficient skill scores for the EMOS$_+$-calibrated forecast using the EMOS$_0$ as the reference forecast. With the exception of a few stations, the EMOS$_+$-calibrated forecast improves significantly (average skill score for the NMAE is 8$\%$, average skill score for the correlation coefficient is 15$\%$).\\
Note that considering as reference the forecast calibrated with the EMOS$_0$ with a rolling training period equal to 40 days (not shown), as proposed by \cite{gneiting2005calibrated}, for 0-24 hour interval forecast, the average skill score for the NMAE is 9$\%$ while the average skill score for the correlation coefficient is 22$\%$ (they are 8$\%$ and 19$\%$, respectively, for the 24-48 hour forecast interval). Larger training sets are thus beneficial in the present study.\\
Figure \ \ref{fig:Epl_vs_Ereferee} is the analogous of Figure\ \ref{fig:Epl_vs_E0} where the skill of our EMOS$_+$ is evaluated against the variant of EMOS$_0$
where all our conditional variables now appear as additional predictors.  
From this figure, a clear added value of our strategy is evident thus confirming the key role of nonlinearities in efficiently disentangling the model error.

We now move to the strategies to deal with the forecasts on the 4 grid points around each station. Two different strategies will be compared. One is simply based on selecting the best grid point based on the correlation coefficient evaluated in the training set. The second method uses the EMOS$_4$ described in section \ref{The EMOS_4 strategy}.

\begin{figure}[h!]
\includegraphics[width=\textwidth]{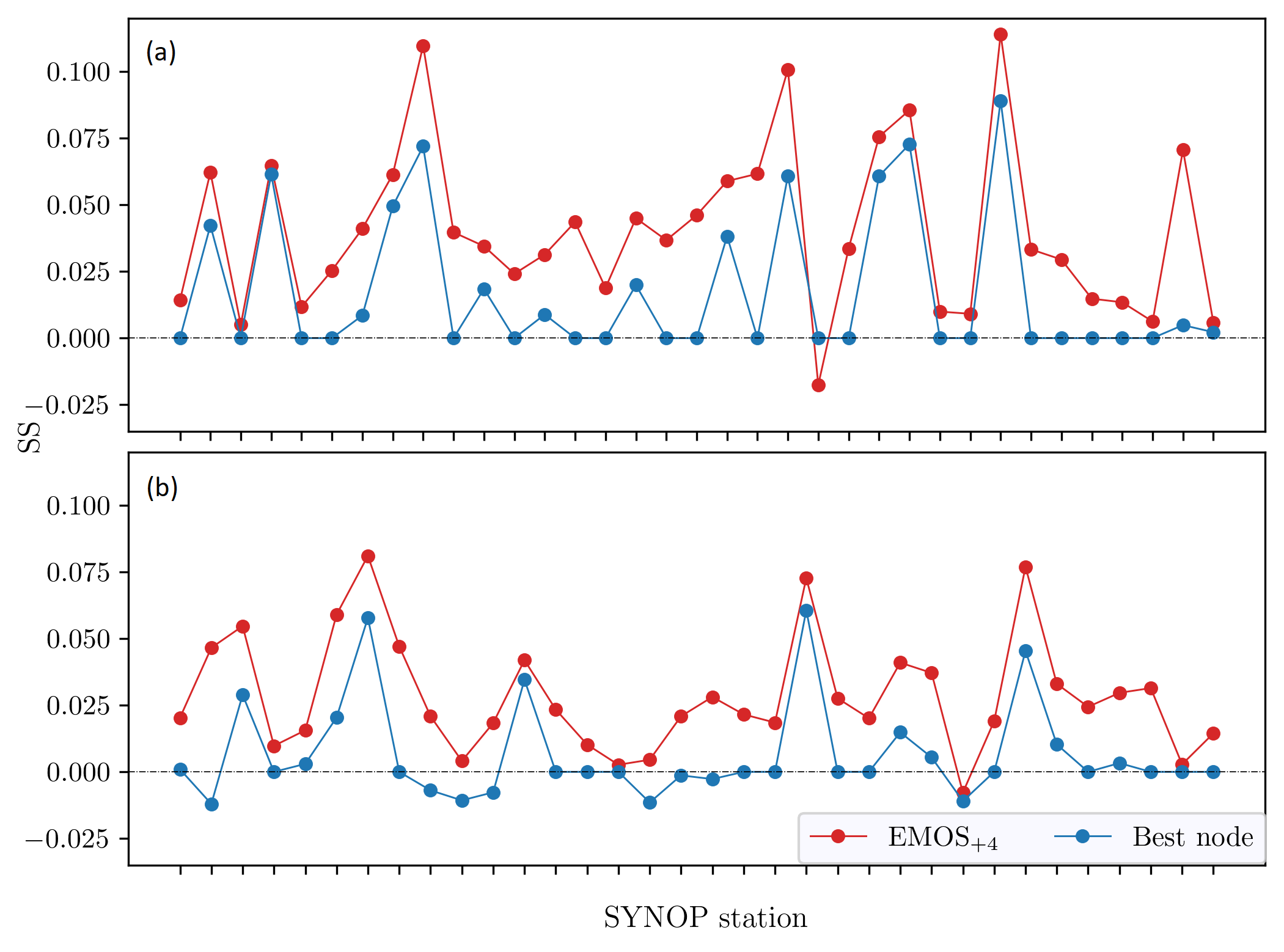}
\caption{Skill score of the NMAE for the forecasts obtained from two different strategies to deal with forecasts on the 4 grid points around a SYNOP station. Results are shown for different SYNOP stations (panel a): stations from 1 to 35; panel b): stations from 36 to 69). Red: EMOS$_{+4}$, blue: the best node evaluated in terms of the highest correlation coefficient. The EMOS$_+$ on the closest-to-the-station model grid point is used as reference.
The average SS are 3$\%$ and 1$\%$ respectively. As one can see, the best strategy is EMOS$_{+4}$. For the 24-48 hour forecast interval (not shown) the average skill score are 3$\%$ (for the NMAE) and 1$\%$ (for the correlation coefficient).}
\label{fig:confronto_E4pl_EMNpl_BAM}
\end{figure}

Figure\ \ref{fig:confronto_E4pl_EMNpl_BAM} shows the NMAE skill score for the two strategies discussed above using the EMOS$_+$ calibration based on the node closest to the SYNOP station as the reference forecast. The strategy of selecting the best grid point is clearly the one giving the poorest result with an overall average improvement of only 1$\%$. EMOS$_{+4}$ leads to a more significant improvement of about 3$\%$\footnote{A similar level of improvement has been obtained (not shown) exploiting a standard Bayesian Model Averaging (BMA) \citep{leith1974theoretical, kass1995bayes, hoeting1999bayesian}.}.\\

\begin{figure}[h!]
\includegraphics[width=\textwidth]{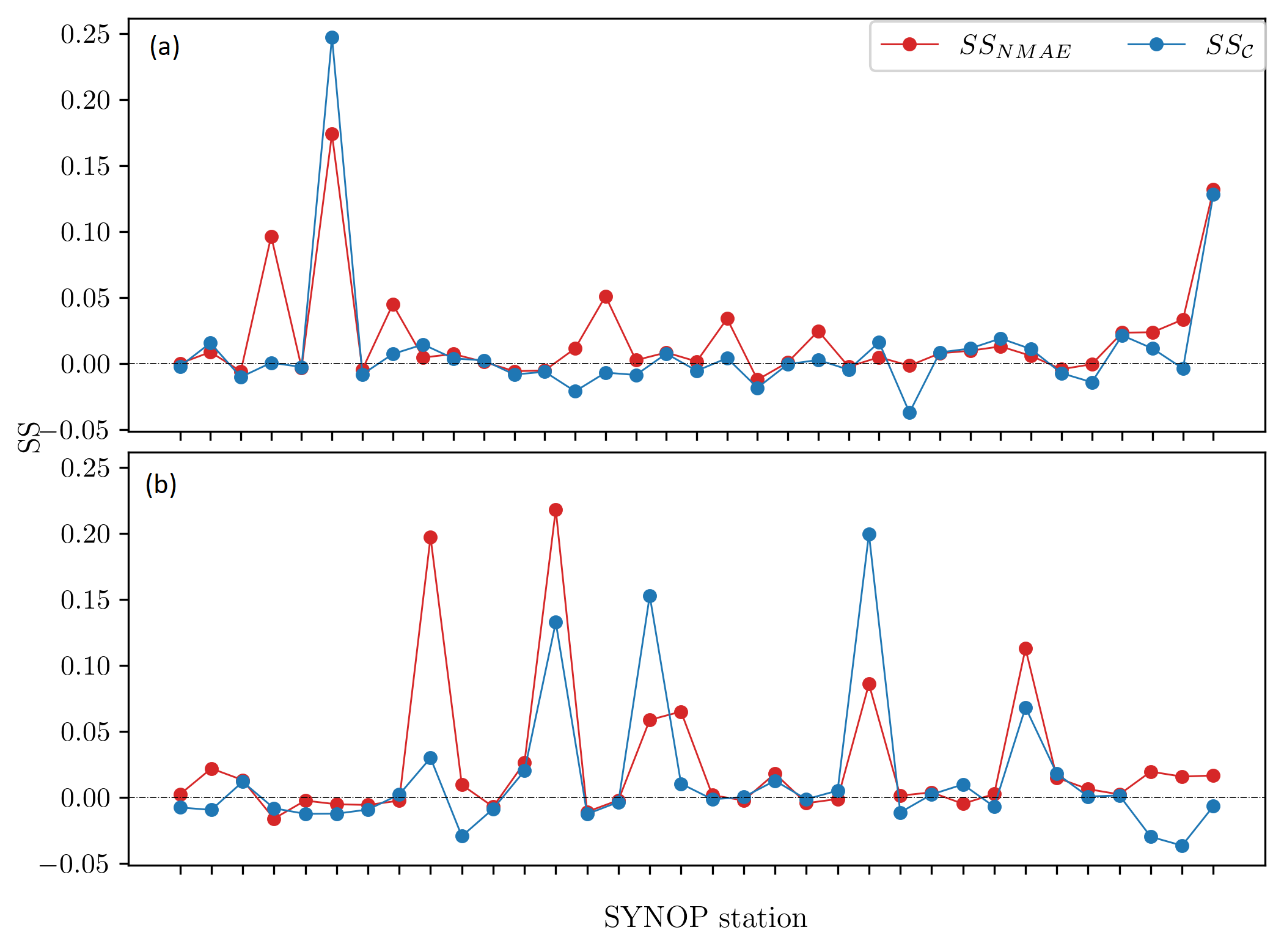}
\caption{Skill score of the forecast from our EMOS$_{+4r}$ calibration using the EMOS$_{+4}$ calibration as the reference forecast for different SYNOP stations (panel (a): stations from 1 to 35; panel (b): stations from 36 to 69). The average skill scores are  2$\%$ (NMAE) and 1$\%$ (correlation coefficient) as for the 24-48 hour forecast interval (not shown).}
\label{fig:NMAE_Pearson_EMOS4plr}
\end{figure}

Let us now assess the added value of our EMOS$_r$-based  calibration downstream of the EMOS$_{+4}$ calibration. Figure \ref{fig:NMAE_Pearson_EMOS4plr} shows the skill scores of both the NMAE and the correlation coefficient of the EMOS$_{+4r}$-based calibration using the EMOS$_{+4}$-based calibrated forecast as reference. The average value of the NMAE skill score is 2$\%$ while the correlation coefficient is 1$\%$. Therefore, the strategy of calibration with the moving training window does not lead, on average, to a great improvement. Nevertheless, some stations clearly show significant improvements while a very small number of stations get a little bit worse.  The effectiveness of the rolling-window training clearly emerges from the reliability index, $\Delta$.
%
\begin{figure}[h!]
\includegraphics[width=\textwidth]{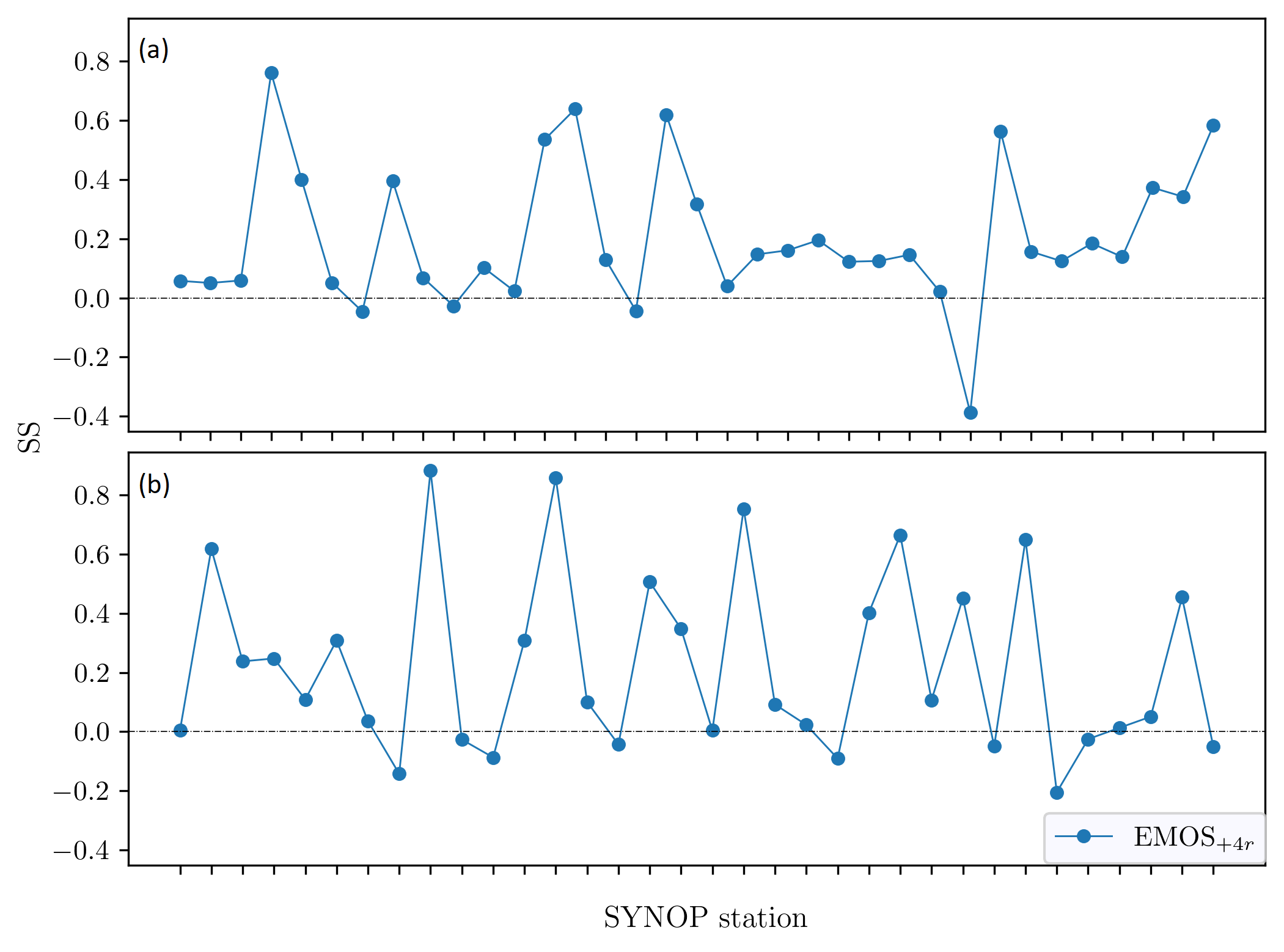}
\caption{$\Delta$ skill score for the EMOS$_{+4r}$ calibration using EMOS$_{+4}$ as the reference prediction for different SYNOP stations (panel (a): stations from 1 to 35; panel (b): stations from 36 to 69). A clear benefit is evident from the rolling strategy: the average SS amounts to the remarkable value of 21$\%$. For the 24-48 hour forecast interval (not shown) the average $\Delta$ skill score is 25$\%$.}
\label{fig:E4pl_vs_E4plrl}
\end{figure}
%
%
Figure\ \ref{fig:E4pl_vs_E4plrl} shows the $\Delta$ skill score for the EMOS$_{+4r}$ forecast with the EMOS$_{+4}$ as the reference forecast. This strategy leads to an average improvement of 21$\%$. For the individual stations, the improvement is very variable; with many stations showing a poor improvement while others reaching a skill score of 80$\%$.

\begin{figure}[h!]
\includegraphics[width=\textwidth]{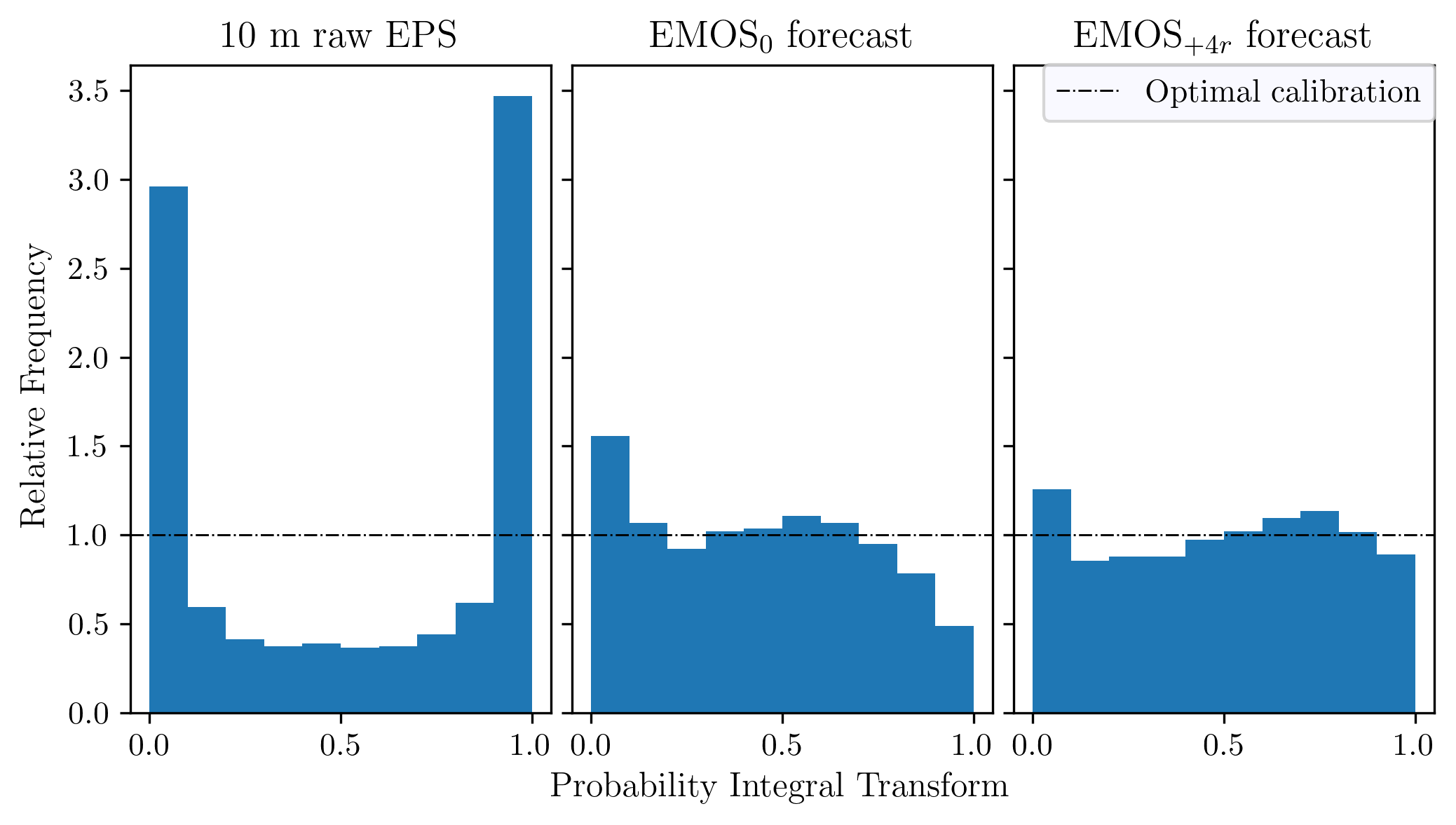}
\caption{PIT histogram of the raw EPS-based forecasts (left panel), the EMOS$_0$-calibrated forecast (center panel), and the EMOS$_{+4r}$ strategy (right panel). From left to right, average values of $\Delta$ are 1.03, 0.18, 0.12, respectively.
The SYNOP station considered here is the one having a value of $\Delta$ (after the EMOS$_{+4r}$ calibration) corresponding to the median (evaluated over all SYNOP stations). For the 24-48 hour forecast interval (not shown), from left to right, average values of $\Delta$ are 0.93, 0.18, 0.12, respectively.}
\label{fig:PIT_prev10_E0_E4plrl}
\end{figure}

Finally, Figure\ \ref{fig:PIT_prev10_E0_E4plrl} shows the evolution of the PIT histogram in three steps of calibration from raw, to EMOS$_0$ and, finally, to EMOS$_{+4r}$ for a SYNOP station having a value of $\Delta$, after the EMOS$_{+4r}$ calibration, corresponding to the median.
As expected, the PIT for the raw forecast shows a strong underdispersion of the variance compared to the actual uncertainty of the model. Also evident is a tendency to underestimate the observations. The EMOS$_0$ calibration has a noticeable effect on the PIT compared to the raw case. EMOS$_{+4r}$ calibration shows a good improvement with an average $\Delta$ passing from 0.18 for the EMOS$_0$-calibrated forecast to 0.12 for both the 0-24 and 24-48 hour forecast interval.

A summary of the quality assessment of our proposed calibrations, from the simplest ones based on observed data to the most complex, is reported in Tab.\ \ref{tab:tabellone_riassuntivo_locale}.

\begin{table}[h!]
\footnotesize
\centering
\caption{Values of the mean (over all 69 SYNOP stations) NMAE, correlation coefficient ($\cal{C}$), and $\Delta$ index for: climatology (Clim); persistence (P24) based on observations at time $t-24$ hours ($t-48$ hours when the look-ahead time is in the 24-48 hour forecast interval);  persistence (P0) defined for all look-ahead times as the observed wind speed at time 00 UTC;
    EMOS$_+$-calibrated P24-based  forecast (EMOS$_{+P24}$);
EMOS$_+$-calibrated P0-based  forecast (EMOS$_{+P0}$);
 10-m wind raw forecast (Raw), EMOS$_0$-based wind forecast (EMOS$_0$), EMOS$_+$-based wind forecast (EMOS$_+$), EMOS$_{+4}$-based wind forecast (EMOS$_{+4}$), and EMOS$_{+4r}$-based wind forecast (EMOS$_{+4r}$).}  
\begin{tabular}{c|ccc|ccc|}
\cline{2-7}
\multirow{2}{*}{}              & \multicolumn{3}{c|}{0-24 hours}                               & \multicolumn{3}{c|}{24-48 hours}                              \\ \cline{2-7} 
                               & \multicolumn{1}{c|}{NMAE} & \multicolumn{1}{c|}{$\cal{C}$}    & $\Delta$ & \multicolumn{1}{c|}{NMAE} & \multicolumn{1}{c|}{$\cal{C}$}    & $\Delta$ \\ \hline
\multicolumn{1}{|c|}{Clim}     & \multicolumn{1}{c|}{0.47} & \multicolumn{1}{c|}{0.31} & 0.16  & \multicolumn{1}{c|}{0.47} & \multicolumn{1}{c|}{0.31} & 0.16  \\ \cline{1-1}
\multicolumn{1}{|c|}{P24}      & \multicolumn{1}{c|}{0.58} & \multicolumn{1}{c|}{0.29} & 0.42  & \multicolumn{1}{c|}{0.63} & \multicolumn{1}{c|}{0.17} & 0.44  \\ \cline{1-1}
\multicolumn{1}{|c|}{P0}       & \multicolumn{1}{c|}{0.54} & \multicolumn{1}{c|}{0.36} & 0.50  & \multicolumn{1}{c|}{0.64} & \multicolumn{1}{c|}{0.11} & 0.53  \\ \cline{1-1}
\multicolumn{1}{|c|}{EMOS$_{+P24}$} & \multicolumn{1}{c|}{0.47} & \multicolumn{1}{c|}{0.35} & 0.14  & \multicolumn{1}{c|}{0.47} & \multicolumn{1}{c|}{0.30} & 0.19  \\ \cline{1-1}
\multicolumn{1}{|c|}{EMOS$_{+P0}$}  & \multicolumn{1}{c|}{0.44} & \multicolumn{1}{c|}{0.46} & 0.14  & \multicolumn{1}{c|}{0.49} & \multicolumn{1}{c|}{0.25} & 0.17  \\ \cline{1-1}
\multicolumn{1}{|c|}{Raw}      & \multicolumn{1}{c|}{0.48} & \multicolumn{1}{c|}{0.64} & 1.03  & \multicolumn{1}{c|}{0.48} & \multicolumn{1}{c|}{0.63} & 0.93  \\ \cline{1-1}
\multicolumn{1}{|c|}{EMOS$_0$}    & \multicolumn{1}{c|}{0.38} & \multicolumn{1}{c|}{0.67} & 0.18  & \multicolumn{1}{c|}{0.38} & \multicolumn{1}{c|}{0.66} & 0.18  \\ \cline{1-1}
\multicolumn{1}{|c|}{EMOS$_+$}    & \multicolumn{1}{c|}{0.35} & \multicolumn{1}{c|}{0.72} & 0.19  & \multicolumn{1}{c|}{0.36} & \multicolumn{1}{c|}{0.71} & 0.18  \\ \cline{1-1}
\multicolumn{1}{|c|}{EMOS$_{+4}$}   & \multicolumn{1}{c|}{0.34} & \multicolumn{1}{c|}{0.74} & 0.18  & \multicolumn{1}{c|}{0.35} & \multicolumn{1}{c|}{0.73} & 0.18  \\ \cline{1-1}
\multicolumn{1}{|c|}{EMOS$_{+4r}$}  & \multicolumn{1}{c|}{0.33} & \multicolumn{1}{c|}{0.75} & 0.12  & \multicolumn{1}{c|}{0.34} & \multicolumn{1}{c|}{0.73} & 0.12  \\ \hline
\end{tabular}
\label{tab:tabellone_riassuntivo_locale}
\end{table}

\section{Power forecast from calibrated wind speed}
\label{Power forecast from calibrated wind speed}

Having small relative errors in forecasting wind speed does not guarantee small relative errors in the wind power forecast because of the well known non linear (cubic in the bulk zone) character of the wind-to-power transfer function.\\
This section addresses the issue of quantifying the benefit on wind power forecast of a calibration strategy giving accurate wind speed forecasting. For this purpose, we associate to each SYNOP station a wind power as that provided by a wind turbine having a cut-in wind speed of 6.8 knots, a rated wind speed of 22.4 knots, and cut-off wind speed of 42.8 knots, with a nominal power of 2 MW. The turbine here considered, for the sake of example, is a Senvion MM100 of which we take the wind-to-power transfer function.\\
The results we are going to present have been obtained by computing the wind power from the mean of the calibrated EPS members, via the wind-to-power transfer function. We also tried another possible option where each wind-speed EPS member is converted into power via transfer function and then averaged over all the resulting wind-power members. Surprisingly, only very minor variations have been observed between the two options to pass from wind speed to wind power.

\begin{figure}[h!]
\includegraphics[width=\textwidth]{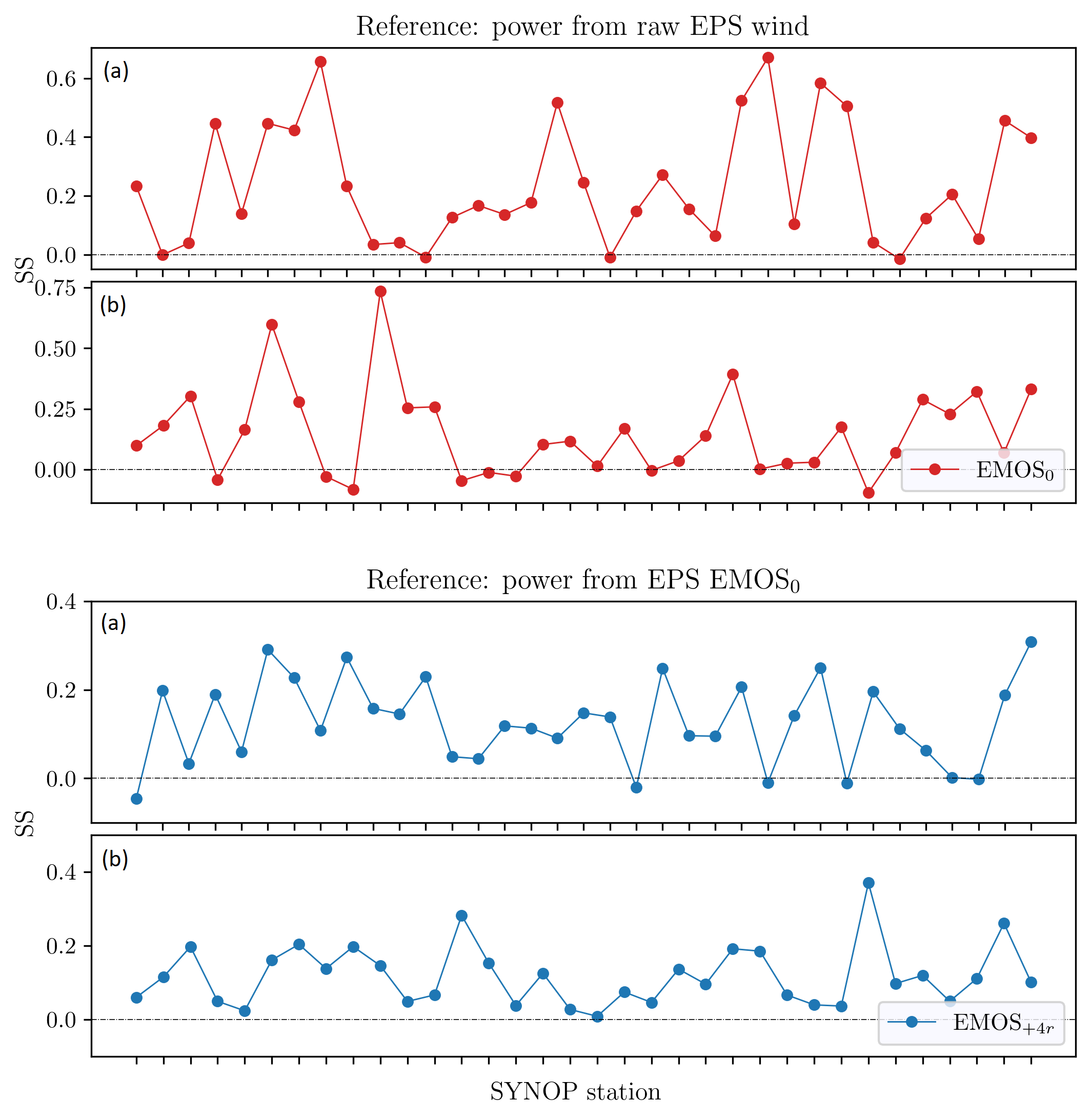}
\caption{NMAE skill score for power forecasts  for different SYNOP stations (panel (a): stations from 1 to 35; panel (b): stations from 36 to 69). Upper panels show the skill score of the power obtained from the EMOS$_0$-calibrated wind speed using the power from the raw EPS wind forecast as reference. Lower panels show the skill score of the power obtained from the EMOS$_{+4r}$ calibrated wind speed using the wind power from the EPS EMOS$_0$-based wind speed forecast as reference. A remarkable improvement can be detected in both cases. On average, the skill score of the two upper panels is 19$\%$, while that of the two lower panels is 12$\%$. For the 24-48 hour forecast interval (not shown) the average skill score of the upper panels is 18$\%$, while that of the lower panels is 10$\%$. The average skill score of EMOS$_{+4r}$ compared to the raw forecast for the 0-24 hour forecast interval is 29$\%$ while for the 24-48 hour interval is 26$\%$ (not shown).}
\label{fig:Emos0_e_Emospl4r_power}
\end{figure}

Figure\ \ref{fig:Emos0_e_Emospl4r_power} shows the NMAE skill score for two different power forecasts. Upper panels shows the NMAE skill score of the power forecast obtained by the wind forecast calibrated with the EMOS$_0$ strategy using as a reference the power forecast obtained from the 10 m raw EPS wind speed forecast. The calibration brings a clear improvement in all stations with an average skill score value of 19$\%$. Lower panels shows the NMAE skill score of the power forecast obtained by the EPS wind forecast calibrated with the EMOS$_{+4r}$ using as a reference the wind power from the EPS EMOS$_0$-based wind speed forecast. Also in this case there is a clear improvement in all stations with an average skill score of 12$\%$. Overall, power obtained from the EMOS$_{+4r}$ wind speed calibration compared to the power forecast from the raw EPS wind speed forecast has an NMAE skill score of 29$\%$.\\
The success of the calibrations shown for the wind speed is thus confirmed also for the wind power prediction.

\begin{figure}[h!]
\includegraphics[width=\textwidth]{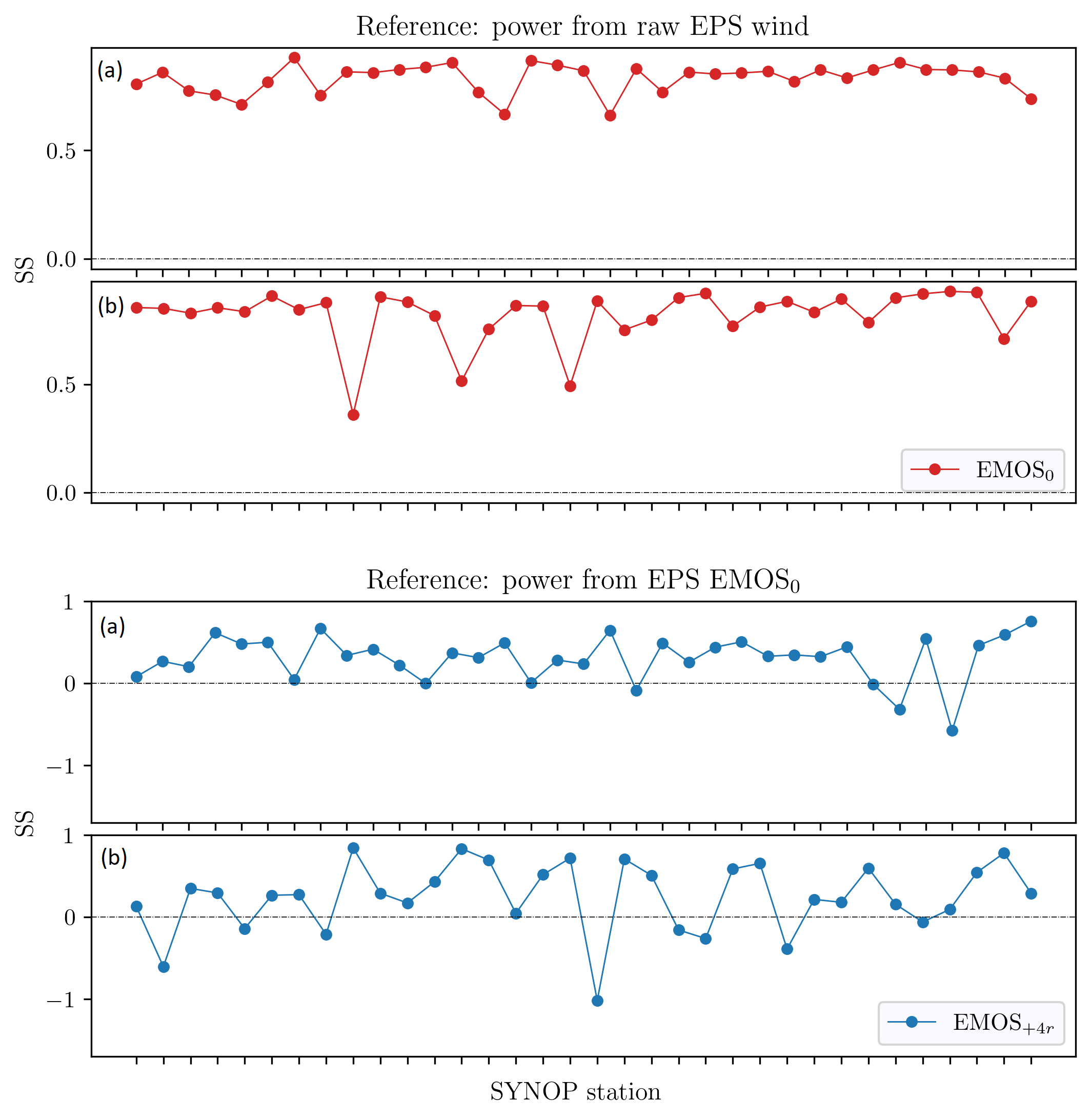}
\caption{$\Delta$ skill score for power forecasts (panel (a): stations from 1 to 35; panel (b): stations from 36 to 69). Upper panels show the skill score of the forecast obtained from the EMOS$_0$-calibrated wind forecast using the power from the raw EPS wind forecast as reference. Lower panels shows the skill score of the power obtained from the EMOS$_{+4r}$-calibrated wind speed using the wind power from the EPS EMOS$_0$-based wind speed forecast as reference. A remarkable improvement can be detected in upper panels while in lower panels the results are more variable. On average, the skill score of upper panels is 83$\%$, while that of lower panels is 29$\%$. For the 24-48 hour forecast interval (not shown) the average skill score of upper panels is 80$\%$, while that of lower panels is 30$\%$.}
\label{fig:delta_EMOSbest_E0_power}
\end{figure}

Let us now pass to analyze the quality of our calibration for the power ensemble forecast as a whole.\\
Figure\ \ref{fig:delta_EMOSbest_E0_power} shows the $\Delta$ skill score for two different cases. The one in the upper panels represents the skill score of the power from the EMOS$_0$-calibrated wind speed forecast taking the power forecast from the raw EPS wind speed as reference. An improvement of 83$\%$ on average is clearly detectable. In the lower panels the skill score of the power forecast obtained from the EMOS$_{+4r}$-calibrated wind speed is shown taking the power forecast from the EMOS$_0$-calibrated EPS wind speed as reference. The improvement is now of about 29$\%$, and it appears more variable from station to station.


A summary of the relevant statistical indices associated to our best calibration strategy is reported in Tabs.\ \ref{tab:tabella_finale} and \ref{tab:tabella_finale_ss}.

\begin{table}[h!]
\footnotesize
\centering
\caption{Values of the mean (over all stations) NMAE, correlation coefficient ($\cal{C}$), and $\Delta$ index, for the raw 10-m wind forecast (Raw) and our best calibration strategy (EMOS$_{+4r}$). Results are reported for the 0-24 and 24-48 hour forecast intervals for both wind and wind power.}  
  \begin{tabular}{c|c|c|c|c|c|c|c|c|}
    \cline{2-9}
    & \multicolumn{4}{c|}{Wind}                              & \multicolumn{4}{c|}{Power}                             \\ \cline{2-9} 
                                & \multicolumn{2}{c|}{0-24 hours} & \multicolumn{2}{c|}{24-48 hours} & \multicolumn{2}{c|}{0-24 hours} & \multicolumn{2}{c|}{24-48 hours} \\ \cline{2-9} 
                                & Raw    & EMOS$_{+4r}$     & Raw    & EMOS$_{+4r}$      & Raw   & EMOS$_{+4r}$      & Raw    & EMOS$_{+4r}$   \\ \hline
\multicolumn{1}{|c|}{NMAE}      & 0.48   & 0.33             & 0.48   & 0.34              & 1.04  & 0.69              & 1.02   & 0.71              \\ \cline{1-1}
\multicolumn{1}{|c|}{$\cal{C}$} & 0.64   & 0.75             & 0.63   & 0.73              & 0.58  & 0.70              & 0.57   & 0.67              \\ \cline{1-1}
\multicolumn{1}{|c|}{$\Delta$}  & 1.03   & 0.12             & 0.93   & 0.12               & 0.98  & 0.08              & 0.70   & 0.08              \\ \hline
  \end{tabular}
\label{tab:tabella_finale}
\end{table}

\begin{table}[h!]
\footnotesize
\centering
\caption{Values of the mean (over all stations) skill scores for both NMAE, correlation coefficient ($\cal{C}$), and $\Delta$ index, of our best calibration (EMOS$_{+4r}$), using raw 10-m wind forecast (Raw) and the EMOS$_0$-based wind calibration forecast (EMOS$_0$) as reference. Results are reported for the 0-24 and 24-48 hour forecast intervals for both wind and wind power.}  

  \begin{tabular}{c|c|c|c|c|c|c|c|c|}
    \cline{2-9}
    & \multicolumn{4}{c|}{EMOS$_{+4r}$-based wind calibration}                              & \multicolumn{4}{c|}{Power from calibrated wind}                             \\ \cline{2-9} 
                                & \multicolumn{2}{c|}{0-24 hours} & \multicolumn{2}{c|}{24-48 hours} & \multicolumn{2}{c|}{0-24 hours} & \multicolumn{2}{c|}{24-48 hours} \\ \cline{2-9} 
                                & Raw    & EMOS$_{0}$     & Raw    & EMOS$_{0}$      & Raw   & EMOS$_{0}$      & Raw    & EMOS$_{0}$   \\ \hline
\multicolumn{1}{|c|}{SS$_{NMAE}$}  & 0.28   & 0.13            & 0.26  & 0.12              & 0.29  & 0.12              & 0.26  & 0.10              \\ \cline{1-1}
\multicolumn{1}{|c|}{SS$_{\cal C}$}  & 0.29   & 0.23             & 0.25   & 0.20              & 0.25  & 0.17             & 0.21   & 0.14              \\ \cline{1-1}
\multicolumn{1}{|c|}{SS$_{\Delta}$}  & 0.89   & 0.21             & 0.89   & 0.22              & 0.87  & 0.27              & 0.88   & 0.29              \\ \hline
  \end{tabular}
\label{tab:tabella_finale_ss}
\end{table}

A final comment on the results reported in Table \ref{tab:tabella_finale_ss} is worth discussing. Our strategy shows indeed remarkably high  mean skill scores for both the NMAE index and the correlation coefficient. The same conclusion holds for both wind speed and wind power.
  These indices are high even when  compared against  other state-of-the-art calibration strategies used for the wind speed. In way of example, with respect to the new, very recently proposed, Machine-Learning-EMOS-based calibration by \cite{baran2021calibration}, we have skill scores at least one order of magnitude larger. Whether or not this might depend also on the type of used datasets (both of the observed data and of the data from the weather prediction models) is an issue to address in future research activities.\\



\clearpage

\section{Conclusions and perspectives}
\label{Conclusions and perspectives}

We have proposed novel EMOS strategies through which nonlinear relationships, between predictands and both predictors and other weather observables used as conditioning variables, can be easily accounted for without having to specify appropriate link functions.\\
On the basis of meteorological observations collected in the years 2018 and 2019 by 69 SYNOP stations over Italy, we have performed a systematic study of different evolutions of known EMOS strategies applied to the Ensemble Prediction System in use at the European Centre for Medium-Range Weather Forecasts.\\
As a first application, we considered observation-driven forecasts: the well-known persistence and climatology-based forecasts. In their basic form (i.e.~without calibration), they are usually exploited as a reference to quantify the added value of more complex calibration strategies. Indeed, the complexity of a strategy must be turned in greater accuracy in order to justify the efforts behind it.\\
From our analysis it turns out that our calibrated persistence outperforms the raw EPS wind forecasts at the SYNOP stations in the 0-24 hour forecast horizon. Climatology overcomes persistence in the 24-48 hour forecast interval and it turned out to be comparable to the raw EPS forecast. These findings are of interest in situations where either no information from NWP models is available or it is available but no calibration has been performed.\\
Moving to calibrations applied to the EPS forecasts, we have considered different possible strategies ordered into a hierarchy of complexity, from the simplest to the most complex. In all cases, a systematic assessment has been performed in order to verify whether a more complex strategy provides a larger added value  than simpler strategies. Our results justify the use of our most elaborated strategy we have called EMOS$_{+4r}$. \\
The following novel ingredients entered into our calibrations: the predictive probability density function entering the EMOS strategy is in our study conditional on several meteorological observables one identifies as suitable to disentangle the structure of the model error. Conditioning variables are a simple way to insert nonlinearities in the ordinary EMOS strategy; the coarse-grained character of NWP models is explicitly taken into account in our calibrations: which node of the computational grid  one has to select as the most representative of the station conditions has been dealt with in a statistically optimal way;\\
Our results pave the way for future applications to wind hindcast calibration, in particular of the 10-m wind speed which is commonly used to force wave models through which evaluate the wave/wind potential in a given sea/land region or to investigate coupling mechanisms occurring at the sea-atmosphere interface \citep{mentaschi2013developing, mentaschi2013problems, mentaschi2015performance, besio2016wave, ferrari2020optimized, rizza2021evaluation, lira2021future}.\\
Our EMOS$_{+4r}$ calibration strategy gave birth to the operative wind forecasting system over Italy consisting of 48-hour look ahead time forecasts accessible on daily basis at the  department web page \citep{WinNT}.

As a last point of our study, we have investigated whether the benefit of a good calibration for the wind speed  forecast brings advantages for the wind power forecast. This issue does not have, a priori, a trivial  answer. Indeed, the cubic relationship between wind speed and wind power makes power forecast even more challenging than wind speed forecast: an acceptable relative error in the wind speed may appreciably downgrade (up to a factor three) in the wind power forecast. The possible downgrade may occur when the wind power is obtained (as done in the present study) from the wind speed via the turbine wind transfer function. As a result of our analysis, it turns out that a net benefit of the calibration carried out for the wind speed forecast also remains for the wind power forecast.\\
This latter result indicates that the continuous effort spent in trying to reduce the forecast error of the wind speed forecast brings its valuable contribution also to the wind power forecast.






\section{Acknowledgments}
\noindent
G.C. has been funded by the Italian banking foundation ``Fondazione Carige''. A.M. acknowledges the funding from the Interreg Italia-Francia Marittimo SICOMAR+ project (grant number D36C17000120006) and from the Compagnia di San Paolo (Project MINIERA No. I34I20000380007). We thank the Aeronautica Militare - Servizio Meteorologico - for providing us with the SYNOP data as well as data from the EPS forecasts. Discussions with Daniele Lagomarsino Oneto, Lorenzo Rosasco, Agnese Seminara, and Alessandro Verri  are warmly acknowledged.

\bibliographystyle{elsarticle-harv} 
\bibliography{biblio.bib}

\end{document}